\documentclass[a4paper,12pt,twoside]{article}
\usepackage{graphicx}
\usepackage{amssymb,amsmath,amsfonts,verbatim,xkeyval,bm}

\def\epsilon{{\varepsilon}}
\def\theta{{\vartheta}}
\def\Ascr{{\cal A}}
\def\Bscr{{\cal B}}
\def\Cscr{{\cal C}}
\def\Escr{{\cal E}}
\def\Gscr{{\cal G}}
\def\Hscr{{\cal H}}
\def\Jscr{{\cal J}}
\def\Kscr{{\cal K}}
\def\Lscr{{\cal L}}
\def\Oscr{{\cal O}}
\def\Rscr{{\cal R}}
\def\Sscr{{\cal S}}
\def\reali{\mathbb{R}}

\def\toro{\mathbb{T}}
\def\vet#1{{\bm #1}}
\def\Lie#1{\Lscr_{#1}}

\title{\bf A numerical criterion evaluating\\ the robustness of planetary\\
   architectures; applications to\\ the $\upsilon$~Andromed{\ae} system}

\author{{\bf Ugo LOCATELLI}$^1$, {\bf Chiara CARACCIOLO}$^2$,
       \\ {\bf Marco SANSOTTERA}$^2$ \& {\bf Mara VOLPI}$^1$\\
{\small $^1$Dipartimento di Matematica dell'Universit\`a degli Studi di
  Roma}\\
{\small ``Tor Vergata'', via della ricerca scientifica 1, 00133 Roma,
  Italy}\\
{\small $^2$ Dipartimento di Matematica dell'Universit\`a degli Studi
  di Milano,}\\
{\small via Saldini 50, 20133 Milano, Italy}\\
{\small emails: {\tt locatell@mat.uniroma2.it, chiara.caracciolo@unimi.it,}}\\
{\small {\tt marco.sansottera@unimi.it, volpi@mat.uniroma2.it}}
}

\date{}

\begin{document}

\maketitle

\begin{abstract} We revisit the problem of the existence of
  KAM tori in extrasolar planetary systems.  Specifically, we consider
  the $\upsilon$~Andromed{\ae} system, by modelling it with a
  three-body problem. This preliminary study allows us to introduce a
  natural way to evaluate the robustness of the planetary orbits,
  which can be very easily implemented in numerical explorations. We
  apply our criterion to the problem of the choice of a suitable
  orbital configuration which exhibits strong stability properties and
  is compatible with the observational data that are available for the
  $\upsilon$~Andromed{\ae} system itself.
\end{abstract}

\bigskip

\markboth{U. Locatelli, C. Caracciolo, M. Sansottera \& M. Volpi}{A
  $\ldots$ criterion evaluating the robustness of planetary
  architectures $\ldots$}

\section{Introduction}
\label{sec:intro}

From the very beginning of their history, physical sciences have been
an inexhaustible source of problems and inspiration for mathematics.
In particular, the orbital characteristics of more and more extrasolar
systems are raising very challenging questions which concern the
modern theory of stability for planetary Hamiltonian systems.

Since the announcement of the discovery of the first one
(see~\cite{May-Que-1995}), thousands of exoplanets have been detected.
Systems hosting more than one planet show a rather surprising variety
of configurations which can be remarkably different with respect to
that of the Solar System, which presents planetary orbits that are
well separated, quasi-circular and nearly coplanar. The situation is
made even more complex by the fact that none of the detection methods
nowadays available to discover extrasolar planets is able to measure
all their orbital elements. In this regard, the Radial Velocity
(hereafter, RV) method is the most effective observation technique,
because it provides values for the semi-major axis $a$, the
eccentricity $e$, and the argument of the pericentre $\omega$ of an
exoplanet (see, e.g.,~\cite{Perryman-2018}). Moreover, the RV method
is able to evaluate the so-called minimal mass $m\sin(\iota)$, where
$m$ and $\iota$ are the mass and the inclination\footnote{More
  precisely, $\iota$ refers to the inclination of the Keplerian
  ellipse with respect to the plane orthogonal to the line of sight
  (i.e., the direction pointing to the object one is observing), which
  is usually said to be ``tangent to the celestial sphere''.} of the
observed exoplanet, respectively. Indeed, this is as a very serious
limitation of the currently available detection techniques, since they
are often unable to completely determine such an important parameter
like the mass of an exoplanet (which is crucial to draw conclusions
about, e.g., its habitability). In particular, the three-dimensional
architecture of a multi-planetary system eludes the observational
measures, when they are made using the RV~method. However, this can be
determined by crossing the results provided by multiple detection
techniques, when different methods can be applied to the same
system. Since the transit photometry is the most prolific technique in
the discovering of exoplanets, its joined use with the RV method is
expected to be very promising for what concerns their orbital
characterisation.  For instance, the combination of the transit and
the RV method allowed to measure the inclination of three exoplanets
orbiting around the L~98-59 star and so to determine rather narrow
ranges for the values of their masses
(see~\cite{Cloutier-et-al-2019}). Although the data obtained through
astrometry are less precise with respect to the aforementioned
detection techniques, they can be joined with the measures provided by
the RV method to evaluate both the inclination $\iota$ and the
longitude of the node $\Omega$ for some massive-enough exoplanets
(e.g., in the case of HD~128311~$c$,
see~\cite{McArthur-et-al-2014}). Moreover, combining astrometry and RV
methods it was possible to determine ranges of values for all the
orbital elements except the mean anomalies $M$ for the two exoplanets
that are expected to be the most massive ones among those orbiting the
$\upsilon$~Andromed{\ae}~A star\footnote{Indeed,
  $\upsilon$~Andromed{\ae} is a binary star. Since the companion is a
  red dwarf that is quite far (about $750$~AU) from the primary star,
  the former is expected to not appreciably affect the planetary
  system orbiting the latter one. For the sake of simplicity, with the
  name of $\upsilon$~Andromed{\ae} hereafter we will refer to both its
  primary star (which is, more precisely, $\upsilon$~Andromed{\ae}~A)
  and the extrasolar system hosting the exoplanets that have been
  discovered around it.} (see~\cite{McArthur-et-al-2010}). On one
hand, this allowed to describe rather carefully the 3D~structure of
the main part of this extrasolar system, with an instantaneous value
of the mutual inclination of $29.9^\circ\pm 1^\circ$; on the other
hand, the uncertainty on the knowledge of a few orbital elements is so
large that the estimated error on the mass of one of the exoplanets is
quite relevant (i.e., $\simeq 30$\%), which is also due to the fact
that its orbital plane is very inclined with respect to the line of
sight.

According to the approach designed in~\cite{Mor-Gio-1995}, the
stability of quasi-integrable systems can be efficiently analysed by
combining the KAM theorem with the Nekhoroshev's one. In fact, their
joint application can ensure the effective stability (that is valid
for interval of times larger than the estimated age of the universe)
for Hamiltonian systems of physical interest.  This strategy has been
successfully applied to a pair of non-trivial planetary models
describing the dynamics of the two or three innermost Jovian planets
of our Solar System; in both those cases, upper bounds on the
diffusion speed have been provided by suitable estimates on the
remainder of the Birkhoff normal form which is preliminarily
constructed in the neighbourhood of an invariant torus
(see~\cite{Gio-Loc-San-2009} \&~\cite{Gio-Loc-San-2017}). The
so-called Arnold diffusion is a phenomenon which cannot take place in
Hamiltonian systems having two degrees of freedom (hereafter, d.o.f.),
because 2D invariant tori act as topological barriers separating the
orbits. Nevertheless, reverse KAM theory can be applied in a way that
is far from being trivial for what concerns the secular dynamics of
extrasolar systems including three bodies (which can be described by a
Hamiltonian model with 2 d.o.f.). In fact, in~\cite{Vol-Loc-San-2018}
the explicit construction of invariant KAM tori is used to infer
information on the possible ranges of values of the mutual
inclinations between the orbital planes of the two exoplanets hosted
in the three following systems: HD~141399, HD~143761 and HD~40307.
However, such an approach suffers serious limitations, mainly due to
the fact that is based on an algorithm which was designed to construct
suitable normal forms for the secular dynamics of our Solar System
(see~\cite{Loc-Gio-2000}). Firstly, this computational procedure is
apparently unable to deal with the case of eccentricities larger
than~$0.1$, which is quite frequent for exoplanets discovered by the
RV detection method. Moreover, although the algorithm constructing the
normal forms can work with bunches of initial conditions at the same
time (if the implementation is made by using interval arithmetics,
see~\cite{Vol-Loc-San-2018}), this kind of procedures can be rather
demanding from a computational point of view, if they are not tailored
carefully to the model under consideration. Therefore, the possibility
to apply extensively such an approach to the study of many extrasolar
systems looks rather doubtful. The so-called criterion of the Angular
Momentum Deficit (hereafter AMD, see~\cite{Las-Pet-2017},
and~\cite{Pet-Las-Bou-2017} for its reformulation adapted to planetary
systems in mean motion resonance) gives an elegant answer to the need
of a ``coarse-graining'' method for quickly studying the stability of
many extrasolar planetary systems.  However, also the AMD criterion
does not cover all the extrasolar planetary systems that are known up
to now, in the sense that is unable to ensure the stability for some
of them. In particular, the AMD criterion can become inapplicable to
systems where the orbital plane of (at least) one exoplanet is highly
inclined with respect to the line of sight; for instance, this is
exactly what occurs in the case of $\upsilon$~Andromed{\ae}, which is
very challenging. On the one hand, the 2D three-body model which
includes the star and its two exoplanets with the largest minimal
masses looks stable according to the AMD criterion, when the line of
sight lies in their common orbital plane (see~Fig.~7
of~\cite{Las-Pet-2017}). On the other hand, when also the inclinations
and the longitudes of the nodes are taken into account, then there is
a remarkable fraction of the possible initial conditions that
generates motions which are evidently unstable
(see~\cite{McArthur-et-al-2010} and~\cite{Deitrick-et-al-2015}). This
is mainly due to the fact that the actual value of the mass of
$\upsilon$~And~\emph{c} should be larger than $5$~times the minimal
one, while the increasing factor affecting the value of
$\upsilon$~And~\emph{d}'s mass is about $2.5\,$.  Therefore, the
perturbation of the Keplerian orbits (that is mainly due to the mutual
gravitation) due to the updated values of the exoplanetary masses is
one order of magnitude larger than the perturbation in the
two-dimensional models of the $\upsilon$~Andromed{\ae} system
considering the data derived by the first observational measures
provided by the RV detection method.

In~\cite{Car-Loc-San-Vol-2022}, we have studied the secular dynamics
of the $\upsilon$~Andromed{\ae} system by adopting the so called
averaged model at order two in the masses. In that framework we have
shown how to construct an invariant (KAM) manifold which is a very
accurate approximation of the orbit originating from initial
conditions that are within the range of the observed values. Moreover,
we have also shown rigorously the existence of such a KAM torus, by
adopting a suitable technique based on a computer-assisted proof. Let
us recall that this ensures that there is a small region around those
initial conditions (and so, consistent with the observational data)
for which the secular dynamics is effectively stable (see again the
aforementioned paper~\cite{Mor-Gio-1995}). Indeed, we have
carefully selected those initial conditions by using a numerical
criterion to evaluate the \emph{robustness} of the corresponding
orbit. The present work is devoted to the description of such a
criterion.  As it will be discussed in the next sections, the concept
of robustness actually refers to the eventually existing torus which
covers the orbit. The key remark which allows us to introduce such a
criterion can be shortly summarised as follows: for what concerns the
secular dynamics of the $\upsilon$~Andromed{\ae} system, a KAM torus
is as more persistent to the perturbing terms as it is closer to a
periodic orbit which corresponds to the anti-alignment of the
pericentre arguments of $\upsilon$~And~\emph{c} and
$\upsilon$~And~\emph{d}.  Thus, it is natural to apply our robustness
criterion in situations where the exoplanets are in a librational
regime with respect to the difference of their pericentre
arguments. Let us recall that $\upsilon$~And~\emph{c} and
$\upsilon$~And~\emph{d} were conjectured to be in such an apsidal
locking state just a few years after their discovery
(see~\cite{Chi-Tab-Tre-2001} and also~\cite{Mich-Mal-2004} for an
explanation of such a dynamical mechanism within the framework of a
secular model). Although our robustness criterion simply applies in
combination with numerical integrations (any averaging procedure is
not strictly necessary), it is somehow related with the dynamical
phenomenon we have described by adopting the language of the normal
forms and the refined computational procedure which is fully detailed
in~\cite{Car-Loc-San-Vol-2022}. Therefore, our robustness numerical
indicator does not aim to be as general as the AMD stability
criterion, at least in its first formulation we are going to
introduce; eventual extensions to contexts different with respect to
the librations in an apsidal locking regime (or in the anti-apsidal
one) could need some nontrivial adaptations.

\section{The orbital dynamics of the exoplanets in the $\upsilon$~Andromed{\ae} system: a short overview}
\label{sec:Ups-Andromedae_overview}

The discovery of three exoplanets orbiting around
$\upsilon$~Andromed{\ae} was made at the end of the last century, by
applying the RV detection method
(see~\cite{Butler-et-al-1999}). Moreover,
in~\cite{McArthur-et-al-2010} it is remarked that a long-period trend
in the analysis of the signals is an indication of the presence of a
fourth planet (named $\upsilon$~And~\emph{e}).  The long-term
stability of a planetary system which includes
$\upsilon$~And~\emph{b}, $\upsilon$~And~\emph{c} and
$\upsilon$~And~\emph{d} has been studied
in~\cite{Deitrick-et-al-2015}, by performing many numerical
integrations; let us also recall that several of them have shown
unstable motions. In the present work, we are going to further
restrict the model by limiting us to consider the two exoplanets that
are expected to be the largest ones. There are good reasons to assume
that the influence exerted by $\upsilon$~And~\emph{b} and
$\upsilon$~And~\emph{e} is negligible: the latter is known quite
poorly (and the RV method is rather sensitive to more massive bodies),
while the former is very tightly close to the star and its minimal
mass is one order of magnitude smaller than the ones of
$\upsilon$~And~\emph{b} and $\upsilon$~And~\emph{c}\footnote{The ratio
  between the semi-major axes of two consecutive planets is $\simeq
  14$ in the case of the pair $\upsilon$~And~\emph{c} --
  $\upsilon$~And~\emph{b}, while it is a bit more than $3$ in the case
  of $\upsilon$~And~\emph{d} -- $\upsilon$~And~\emph{c}. In the case
  of $\upsilon$~And~\emph{b} the value of the quantity
  $m\sin(\iota(0))$ is known to be $0.0594 \pm 0.0003$~$M_J$ (see
  Table~13 of \cite{McArthur-et-al-2010}, whose data concerning
  $\upsilon$~And~\emph{c} and $\upsilon$~And~\emph{d} are reported in
  our Table~\ref{tab:orbel-with-errors} above). Let us also recall
  that the initial inclination $\iota(0)$ of $\upsilon$~And~\emph{b}
  is unknown and, thus, the minimal value is the only information
  available about its mass.}.

\begin{table}
  \begin{center}
    \caption{Orbital elements and minimal masses of the exoplanets
      $\upsilon$~And~\emph{c} and $\upsilon$~And~\emph{d}. All the
      data appearing in the following first three columns are reported
      from Table~13 of \cite{McArthur-et-al-2010}. In the rightmost
      column we have included also the relative errors for each
      quantity.  In all our numerical integrations the stellar mass of
      $\upsilon$~Andromed{\ae} is assumed to be
      $m_0=1.31\,M_{\odot}\,$. As usual, $M_{\odot}$ and $M_J$ denote
      the solar mass and the Jupiter one, respectively.}
  \label{tab:orbel-with-errors}
  \begin{tabular}{l c c c}
    \hline
    & $\upsilon$~And~\emph{c} & $\upsilon$~And~\emph{d} & rel. err. \\ 
    \hline
    $a(0)$ [AU] & $0.829 \pm 0.043$ & $2.53 \pm 0.014$ & $\simeq\,5\,$\% \\
    $e(0)$ & $0.245 \pm 0.006$ & $0.316 \pm 0.006$ &$\simeq\,5\,$\% \\
    $\iota(0)$ [$^\circ$]& $7.868 \pm 1.003$ & $23.758 \pm 1.316$ & $1.4/180$ \\ 
    $\omega(0)$ [$^\circ$] & $247.66 \pm 1.76$ & $252.99 \pm 1.31$ & $1.8/360$\\
    $\Omega(0)$ [$^\circ$] & $236.85 \pm 7.53$ & $4.07 \pm 3.31$ & $7.6/360$  \\
    $m\sin(\iota(0))$ [$M_J$] & $1.96 \pm 0.05$ & $4.33 \pm 0.11$
    & $\lesssim\,3\,$\% \\
    \hline
  \end{tabular}
  \end{center}
\end{table}

The initial values of the orbital elements (except the mean anomalies,
that are unknown) for the pair of exoplanets $\upsilon$~And~\emph{c},
$\upsilon$~And~\emph{d} and their minimal masses are reported in
Table~\ref{tab:orbel-with-errors}. Hereafter, in our three-body model
of the $\upsilon$~Andromed{\ae} extrasolar system the indexes $1,\,2$
will be used to refer to the inner planet and the outer one,
respectively, while $m_0$ will denote the stellar mass. Looking at
Table~\ref{tab:orbel-with-errors}, one can appreciate that all the
reported data are given with a relative uncertainty that is not larger
than a few percentage units. Due to the occurrence of the increasing
factor $1 / \sin(\iota_j(0))$ (with $j=1,\,2$), this is no more true
for the exoplanetary masses. A straightforward evaluation starting
from the data reported in Table~\ref{tab:orbel-with-errors} gives
$m_1=14.6 \pm 2.2$ and $m_2=10.8 \pm 0.9$. Therefore, the relative
uncertainty of at least one parameter (which plays a crucial role in
the discussion about the stability of this extrasolar planetary
system) can reach\footnote{Taking into account all the uncertainties
  due to the observational measures, the errors ranges are even
  wider. Indeed, in Table~13 of \cite{McArthur-et-al-2010} the
  following values for the exoplanetary masses are given:
  $m_1=13.98_{-5.3}^{+2.3}$ and $m_2=10.25_{-3.3}^{+0.7}$.} $15\,$\%
of the corresponding mid value; this is the case of the mass of
$\upsilon$~And~\emph{c}.

We emphasise that an extensive study of the possible motions with an
homogeneous and accurate covering of all the possible initial
conditions and parameters gets immediately far too complex from a
computational point of view, because it would require to deal with a
fourteen-dimensional grid. For the sake of simplicity, we started to
reduce the complexity by fixing some of the parameters that are
determined rather precisely by the observational measures made by
using the RV method; in detail, they are two pairs of orbital
elements, i.e., the initial values of semi-major axes and
eccentricities, and the minimal masses. The latter two pairs have been
fixed so to be equal to the lowest possible values of the
corresponding ranges given in Table~\ref{tab:orbel-with-errors}. This
choice has been made in order to increase the fraction of the orbital
motions that are apparently stable. All the values of the parameters
that have been so fixed by us are reported in
Table~\ref{tab:orbel-partial}.

\begin{table}
  \begin{center}
    \caption{List of the values of the parameters that are kept fixed
      in all our numerical explorations. In the first two rows, the
      initial conditions concerning with semi-major axes and
      eccentricities of the exoplanets $\upsilon$~And~\emph{c} and
      $\upsilon$~And~\emph{d} are reported. In the last row, their
      values of the minimal masses are given.}
  \label{tab:orbel-partial}
  \begin{tabular}{l c c}
    \hline
    & $\upsilon$~And~\emph{c} & $\upsilon$~And~\emph{d} \\ 
    \hline
    $a(0)$ [AU] & $0.829$ & $2.53$ \\
    $e(0)$ & $0.239$ & $0.310$ \\
    $m\sin(\iota(0))$ [$M_J$] & $1.91$ & $4.22$ \\
    \hline
  \end{tabular}
  \end{center}
\end{table}

  \begin{figure}
    \vspace*{-0.6 cm}
    \centering
      \includegraphics[clip, height=4.6cm,]{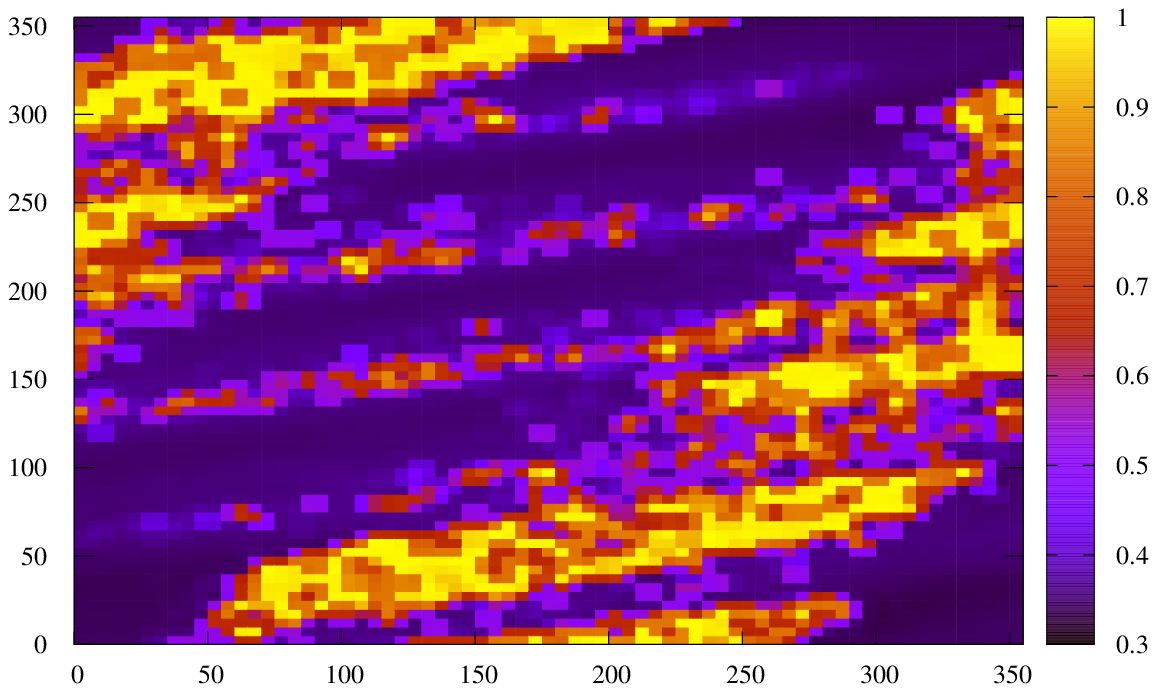}
      \includegraphics[clip, height=4.6cm,]{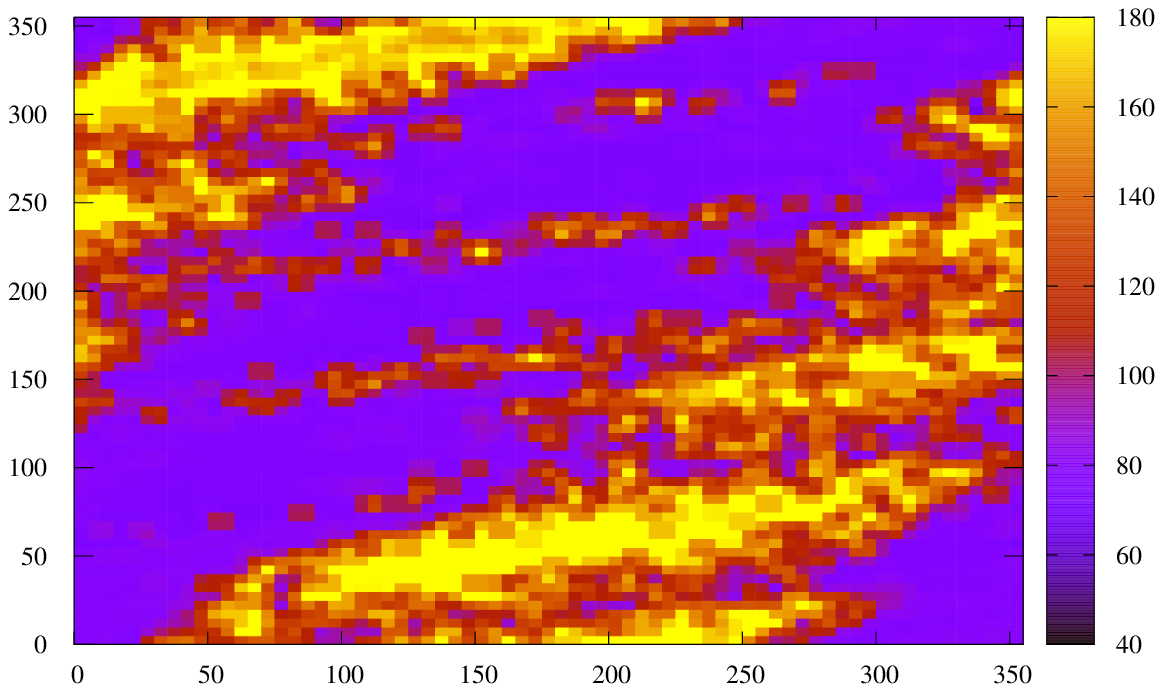}
      \caption{On the left, a colour-grid plot of the maximal value
        reached by the eccentricity $e_2$ of
        $\upsilon$~And~\emph{d}. On the right, the same for the
        maximal value of the difference between the pericentres, i.e.,
        $\max_t\,|\omega_1(t)-\omega_2(t)|$. In both panels, the plots
        are made as a function of the initial values of the mean
        anomalies $M_1(0)$ and $M_2(0)$. See the text for more details
        about the choice of the initial conditions.}
      \label{maxecc_maxdeltaomega}
  \end{figure}

We start our numerical explorations by investigating the dependence on
the pair of the orbital elements that are unknown, namely the mean
anomalies $M_1$ and $M_2\,$\footnote{The observational data reported
  in both online catalogues and published papers determine values for
  the orbital period and the epoch of periastron. From these two
  values it is possible to infer the values of the mean anomalies, but
  they are never explicitly determined. In this respect, we then
  consider them as \emph{unknown}. }. For this purpose, we decide to
consider sets of initial conditions such that $\iota_j(0)$,
$\omega_j(0)$ and $\Omega_j(0)$ are set to be equal to the
corresponding mid values reported in
Table~\ref{tab:orbel-with-errors}, $\forall\ j=1,2$. Moreover, the
initial conditions are complemented with the data reported in
Table~\ref{tab:orbel-partial}, while the initial values of the mean
anomalies are taken from a regular 2D grid covering all the set
$[0^\circ,360^\circ]\times[0^\circ,360^\circ]$ with a grid-step of
$5^\circ$. Hereafter, the mass of each exoplanet is always determined
by multiplying its minimal value (appearing in
Table~\ref{tab:orbel-partial}) by the corresponding increasing factor
$1/\sin(\iota(0))$. Starting from each of the initial conditions
defined just above, we have numerically integrated the Hamilton
equations describing our three-body planetary model, by using the
symplectic method $\Sscr\Bscr\Ascr\Bscr_{\Cscr 3}$ as it is defined
in~\cite{Las-Rob-01} for a timespan of $10^5$~yr, with an integration
step of $0.02$~yr. The main results so obtained are summarised in
Fig.~\ref{maxecc_maxdeltaomega}, which highlights that the choice of
the initial values of the mean anomalies affects the orbital dynamics
in a very remarkable way. In fact, the regions that appear with
lighter colours in the left panel correspond to motions that can
experience close encounters. Let us recall that the threshold value of
the eccentricity of the outer planet on top of which collisions with
the inner planet are possible can be roughly evaluated as
$1-(a_1(0)/a_2(0))\simeq 0.67\,$. On the other hand, about $50\,$\% of
the colour-grid plot in the left panel is in dark; this means that the
maximum value of the eccentricity of the outer planet looks to be
safely below that threshold allowing close encounters with the inner
one. The strong similarity between the two panels of
Fig.~\ref{maxecc_maxdeltaomega} clearly suggests that stable
configurations are possible when the difference of the pericentre
arguments is in a librational regime, i.e., the orbital motions are
such that the maximum of the half-width of the oscillations concerning
with $\omega_1(t)-\omega_2(t)$ is less than $180^\circ$. Let us
emphasise that this kind of phenomena has already been observed in the
last few years. In fact, the relevance of the impact due to the mean
anomalies on the orbital dynamics of extrasolar systems that are close
or in mean motion resonance has been shown, e.g.,
in~\cite{Lib-San-2013} and~\cite{San-Lib-2019}.

\begin{figure}
\begin{center}
 \includegraphics[width=12.4cm]{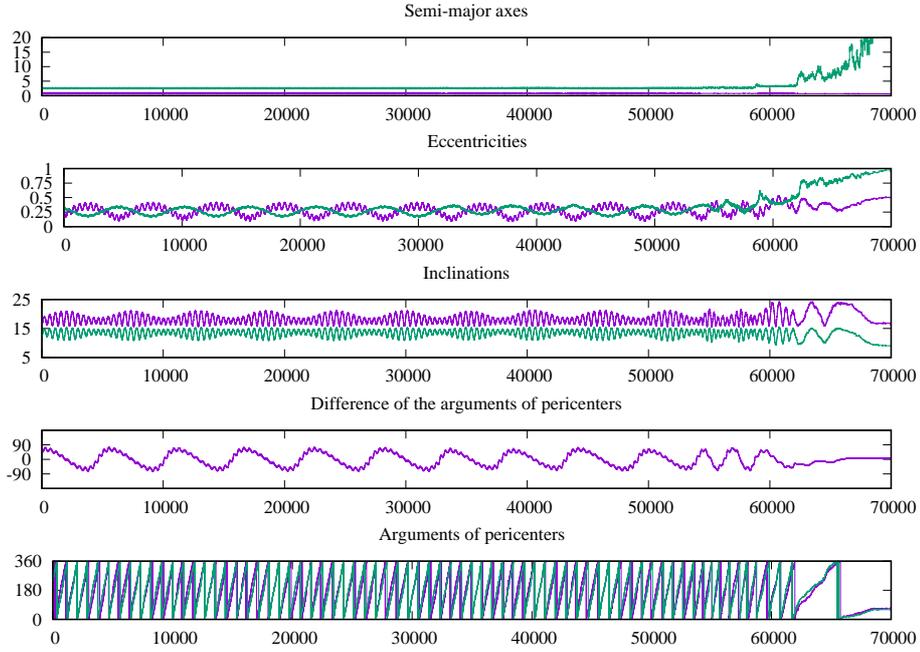} 
 \caption{Orbital evolution of the exoplanets $\upsilon$~And~\emph{c}
   and $\upsilon$~And~\emph{d} in the case of one single set of
   initial conditions among those that have been considered also in
   Fig.~\ref{maxecc_maxdeltaomega} with the particular choice
   $M_1(0)=0^\circ$ and $M_2(0)=120^\circ$ for what concerns the
   initial values of the mean anomalies. From top to bottom, the five
   graphs include the plots of the
   evolution for the following quantities:
   semi-major axes $a_1$ and $a_2\,$, eccentricities $e_1$ and
   $e_2\,$, inclinations $\iota_1$ and $\iota_2\,$, difference of the
   arguments of the pericentres $\omega_2-\omega_1$, arguments of the
   pericentres $\omega_1$ and $\omega_2\,$. The plots in green refer
   to the orbital motion of $\upsilon$~And~\emph{d}. The inclinations
   are evaluated with respect to the direction of the total angular
   momentum. In all the abscissas, the year is assumed as unit of
   measure of time.}
   \label{orbU}
\end{center}
\end{figure}

In order to make clear the ideas, it can be convenient to have a close
look at the dynamical evolution of most of the orbital elements, for a
motion starting form one single set of initial conditions, that is
selected among those considered in Fig.~\ref{maxecc_maxdeltaomega}. In
particular, the orbital evolution described by the plots included in
Fig.~\ref{orbU} refers to $M_1(0)=0^\circ$ and $M_2(0)=120^\circ$; let
us recall that the other values of the initial conditions are taken
from the mid values of Table~\ref{tab:orbel-with-errors} (for what
concerns $\iota(0)$, $\omega(0)$ and $\Omega(0)$ only) and from
Table~\ref{tab:orbel-partial} (for the remaining data).  Looking at
the plots of the orbital elements one can appreciate that the orbit is
unstable; for any of those plots, the lack of quasi-periodicity is
particularly evident after $50000\,$yr. From the behaviour of the
semi-major axes and the eccentricities, it is obvious that the outer
planet is ejected from the system at the end of the numerical
simulation. Let us also stress that our standard implementation of the
symplectic method $\Sscr\Bscr\Ascr\Bscr_{\Cscr 3}$ usually crashes for
all the motions starting from initial conditions which correspond to
regions of lighter colour in the plot of the left panel in
Fig.~\ref{maxecc_maxdeltaomega}.

Hereafter, we will refer to the three-body planetary problem that has
been described in the present section as the \emph{complete} model, in
order to distinguish it with respect to the \emph{secular} one. The
latter is an Hamiltonian system which is defined by a suitable
procedure of averaging that will be briefly discussed in the next
section.

\section{The construction of invariant tori in the secular dynamics of the $\upsilon$~Andromed{\ae} system as a source of inspiration}
\label{sec:Ups-Andromedae_sec_dyn}

The present section is devoted to recall some of the ideas we recently
used in order to successfully construct KAM tori, that are invariant
for the secular dynamics of the $\upsilon$~Andromed{\ae} planetary
system and are also in librational regime with respect to the
difference of the pericentre arguments
(see~\cite{Car-Loc-San-Vol-2022}). Our aim is to explain in a rather
natural way the reasons to introduce our numerical criterion
evaluating the robustness of planetary configurations, that will be
properly defined in the next section.

In the case of the secular dynamics of the $\upsilon$~Andromed{\ae}
planetary system, the preliminary construction of the normal form for
a particular elliptic torus is essential to be performed before the
one constructing the final KAM torus. These two constructive
procedures can be described in an unified way, as we explained
in~\cite{Loc-Car-San-Vol-2022}.  We defer the reader to those
pedagogical notes for all the details about this kind of (so-called)
semi-analytic algorithms, that can be summarised as follows for our
goals.

The proof scheme of the KAM theorem can be formulated in terms of a
constructive algorithm whose convergence is ensured if some suitable
hypotheses are satisfied. This procedure starts by considering an
analytic Hamiltonian function
$H^{(0)}:\Ascr\times\toro^n\mapsto\reali$ (being
$\Ascr\subseteq\reali^n$ an open set) of the form
$H^{(0)}(\vet{p},\vet{q})=\vet{\nu}\cdot\vet{p}+
h^{(0)}(\vet{p},\vet{q})+\epsilon f^{(0)}(\vet{p},\vet{q})$, where $n$
denotes the number of degrees of freedom, $\vet{\nu}\in\reali^n$ is an
angular velocity vector and $h^{(0)}$ is at least quadratic with
respect to the actions $\vet{p}$, i.e.,
$h^{(0)}(\vet{p})=\Oscr(\|\vet{p}\|^2)$ for $\vet{p}\to\vet{0}$.  The
term $\epsilon f^{(0)}(\vet{p},\vet{q})$ appearing in $H^{(0)}$ is
usually called the perturbing term and it is made smaller and smaller
by the normalisation procedure, which is defined by an infinite
sequence of canonical transformations. This entails that we have to
introduce a sequence of Hamiltonians $H^{(r)}$ that are iteratively
defined so that
\begin{equation}
  H^{(r)}=\exp\big(\Lie{\chi_2^{(r)}}\big)\exp\big(\Lie{\chi_1^{(r)}}\big)H^{(r-1)}
  \qquad
  \forall\ r\ge 1
  \ ,
\label{eq:r-th-normalization-step}
\end{equation}
where the generating functions $\chi_1^{(r)}$ and $\chi_2^{(r)}$ are
determined in such a way to remove the part of the perturbation term
that is both $\Oscr(\epsilon^r)$ and not dependent on $\vet{p}$ or
linear in $\vet{p}$, respectively. We then say that
formula~\eqref{eq:r-th-normalization-step} defines the $r$-th
normalization step. We stress that the Lie series operators
$\exp\big(\Lie{\chi_2^{(r)}}\big)$ and
$\exp\big(\Lie{\chi_1^{(r)}}\big)$ define canonical transformations
when they are applied to the whole set of variables
$(\vet{p},\vet{q})$. This is due to the fact that they are given in
terms of the Lie derivatives $\Lie{\chi_2^{(r)}}$ $\Lie{\chi_1^{(r)}}$
(which in turn are expressed as Poisson brackets, i.e.,
$\Lie{g}f=\{f,g\}$ for any pair of dynamical functions $f$ and $g$
that are defined on the phase space). The statement of the KAM theorem
(see~\cite{Kolmogorov-1954}, \cite{Arnold-1963} and~\cite{Moser-1962})
can be shortly formulated as follows:

\begin{description}
  \item
    \emph{if $\vet{\nu}$ is non-resonant enough, $h^{(0)}$ is
      non-degenerate with respect to the actions $\vet{p}$ and the
      parameter $\epsilon$ is small enough, then there is a canonical
      transformation $(\vet{p},\vet{q})=\Psi(\vet{P},\vet{Q})$,
      leading $H^{(0)}$ in the so called Kolmogorov normal form}
\end{description}
\begin{equation}
  \Kscr(\vet{P},\vet{Q})=\nu\cdot\vet{P}+\Oscr(\|\vet{P}\|^2)\ ,
  \label{eq:Kolmogorov-normal-form}
\end{equation}
\emph{being $\Kscr=H\circ\Psi\,$.}

Indeed, the final canonical transformation $\Psi$ is obtained by
composing all the canonical transformations induced by
$\exp\big(\Lie{\chi_1^{(1)}}\big)$,
$\exp\big(\Lie{\chi_2^{(1)}}\big)$,
$\ldots$~$\exp\big(\Lie{\chi_1^{(r)}}\big)$,
$\exp\big(\Lie{\chi_2^{(r)}}\big)$~$\ldots$ Moreover, one can easily
verify that the quasi-periodic motion law
$t\mapsto(\vet{P}(t)=\vet{0}\,,\,\vet{Q}(t)=\vet{Q}_0+\vet{\nu}t)$ is
the unique solution for the Hamilton equations related to the
Kolmogorov normal form~\eqref{eq:Kolmogorov-normal-form} with initial
conditions $(\vet{P}(0)\,,\,\vet{Q}(0))=(\vet{0}\,,\,\vet{Q}_0)$.
Since the canonical transformations have the property of preserving
solutions, then the $n$-dimensional KAM torus
$\big\{(\vet{p},\vet{q})=\Psi(\vet{0},\vet{Q})\,,
\>\forall\>\vet{Q}\in\toro^n\big\}$ is invariant with respect the flow
induced by the initial Hamiltonian $H^{(0)}$.

\subsection{Preliminaries}\label{sbs:prel}
As it has been first explained in~\cite{Loc-Gio-2000}, the so-called
secular model at order two in the masses can be properly introduced by
performing a first step of normalization, which aims at removing the
perturbation terms depending on the fast revolution angles. In order
to set the ideas let us recall that a three-body Hamiltonian problem
has nine degrees of freedom. Three of them can be easily separated
because they describe the uniform motion of the centre of mass in an
inertial frame. The nontrivial part of the dynamics is represented in
astrocentric canonical coordinates and its degrees of freedom can be
further reduced by two using the conservation of the total angular
momentum $\vet C$. As it is shown in section~6 of~\cite{Laskar-1989},
this allows us to write the Hamiltonian as a function of four pairs of
Poincar\'e canonical variables, that are
\begin{equation}
  \label{frm:poin-var}
  \vcenter{\openup1\jot 
    \halign{
      $\displaystyle\hfil#$&$\displaystyle{}#\hfil$&$\displaystyle#\hfil$
      &$\displaystyle\hfil#$&$\displaystyle{}#\hfil$&$\displaystyle#\hfil$\cr
      \Lambda_j &= &\frac{m_0 m_j\sqrt{G(m_0+m_j)a_j}}{m_0 + m_j}\,,
      \ &\xi_j & = &\sqrt{2 \Lambda_j}
      \sqrt{1-\sqrt{1-e_j^2}}\cos{(\omega_j)}\,,
      \ \cr
      \lambda_j &= & M_j + \omega_j\,,
      &\eta_j &= &-\sqrt{2 \Lambda_j} \sqrt{1-\sqrt{1-e_j^2}}\sin{(\omega_j)}
      \,,
      \cr
  }}
  \ \forall \> j=1,2.
\end{equation}
We also recall that the reduction of the total angular momentum makes
implicit the dependence on the orbital elements that are missing in
formula~\eqref{frm:poin-var}. They are the inclinations and the
longitudes of the nodes, which are conveniently expressed with respect
to the so-called Laplace invariant plane, that is orthogonal to the
total angular momentum $\vet C$. However, also the instantaneous
values of these two pairs of orbital elements can be recovered by the
knowledge of all the others and the euclidean norm of $\vet C$.  The
actions $\Lambda_1$ and $\Lambda_2$ (that are conjugate with respect
to the mean anomalies $\lambda_1$ and $\lambda_2\,$, respectively) are
usually expanded around a pair of reference values, namely
$\Lambda_1^*$ and $\Lambda_2^*$. These values are obtained by replacing the
semi-major axes appearing in the corresponding definition included in
formula~\eqref{frm:poin-var} with their initial values $a_1(0)$ and
$a_2(0)$ reported in Table~\ref{tab:orbel-partial}. Thus, after the
reduction of the constants of motion, the Hamiltonian describing the
three-body planetary problem can be expressed as a function of four
pairs of canonical variables: $\vet{L}=\vet{\Lambda}-\vet{\Lambda}^*$,
$\lambda$, $\xi$ and $\eta$.  We can introduce the secular model at
order two in the masses thanks to the following three operations: we
perform a first step of normalization aiming to reduce the perturbing
part that does not depend on $\vet{L}$ and does depend on the angles
$\lambda$; we put $\vet{L}=\vet{0}$ (this is made because we expect
that the oscillations of the semi-major axes close to their initial
values have negligible effects); we finally average over the mean
anomalies $\vet{\lambda}$ (as it is usual, when the analysis is
focused on the long-term evolution of a planetary system).
Therefore, we can write our secular Hamiltonian model as follows:
\begin{equation}
\label{frm:hsec}
H^{({\rm sec})}(\vet \xi,\vet \eta) =
\sum_{s=1}^{N_S/2} h^{({\rm sec})}_{2s}(\vet \xi,\vet \eta)\ ,
\end{equation}
where $h_{2s}$ is an homogeneous polynomial of degree $2s$. This means
that the expansion contains just terms of even degree, as a further
consequence of the well known D'Alembert rules. Let us stress that the
canonical variables $(\vet \xi,\vet \eta)$ appearing in
formula~\eqref{frm:hsec} are not the ones defined
in~\eqref{frm:poin-var}, by abuse of notation. Indeed, the former
variables are obtained from the latter ones, by performing the
canonical transformation defined by the normalization step introducing
the secular model at order two in the masses. Since this change of
variables differs from the identity, because of a small correction
that is of order one in the masses, then the values of the canonical
variables $(\vet \xi,\vet \eta)$ appearing in formula~\eqref{frm:hsec}
are quite close to the corresponding ones that are defined
in~\eqref{frm:poin-var}. These last comments joined with the remark
that both $\xi_j$ and $\eta_j$ are $\Oscr(e_j)$ for $e_j\to 0$
$\forall\>j=1,2$ (as it can be easily checked by looking at the
definition~\eqref{frm:poin-var}) allow us to give a meaning to the
parameter $N_S\,$, in the sense that $H^{({\rm sec})}$ provides an
approximation of the secular dynamics up to order $N_S$ in the
eccentricities. On the one hand, in practical applications one is
interested in expansions up to high order in
eccentricities\footnote{However, it must be taken into account that
  too large expansions of the secular model introduced here can be
  meaningless, because the high quality of the approximation in the
  eccentricities can be shadowed by the lack of precision with respect
  to the masses.}; on the other hand, the computational effort
critically increases with respect to $N_S\,$.  To fix the ideas, in
the case of the $\upsilon$~Andromed{\ae} planetary system we have
found that setting $N_S=8$ is a good balance between these two
different needs that are in opposition to each other.

We have explicitly performed all the computations of Poisson brackets
(required by Lie series formalism to express canonical
transformations) and all the expansions briefly described in the
present section, by using {\it X$\rho$\'o$\nu o\varsigma$}. It is a
software package especially designed for doing computer algebra
manipulations into the framework of Hamiltonian perturbation theory
(see~\cite{Gio-San-Chronos-2012} for an introduction to its main
concepts). Such computations also allow an easy visualisation of the
secular dynamics by adopting a classical tool in the context of the
numerical investigations: the Poincar\'e sections.  In fact, we have
performed many numerical integrations of the secular model $H^{({\rm
    sec})}$ that is defined in~\eqref{frm:hsec} by simply applying the
RK4 method\footnote{It is very well known that long-term numerical
  integrations of secular models are much less computationally
  expensive than those dealing with the corresponding complete
  planetary system (see, e.g.,~\cite{Laskar-1988} and the references
  therein).}.

  \begin{figure}
    \centering
      \includegraphics[clip, height=4.6cm,]{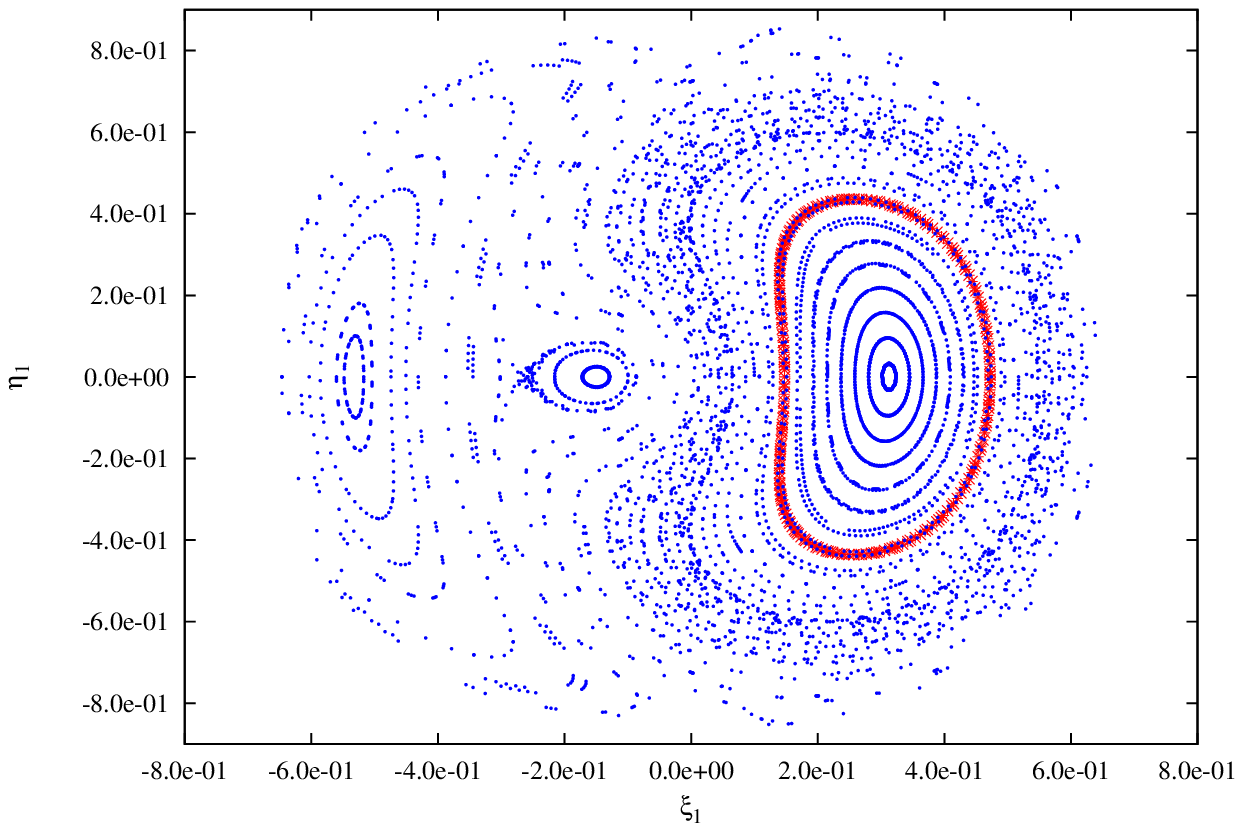}
      \includegraphics[clip, height=4.6cm,]{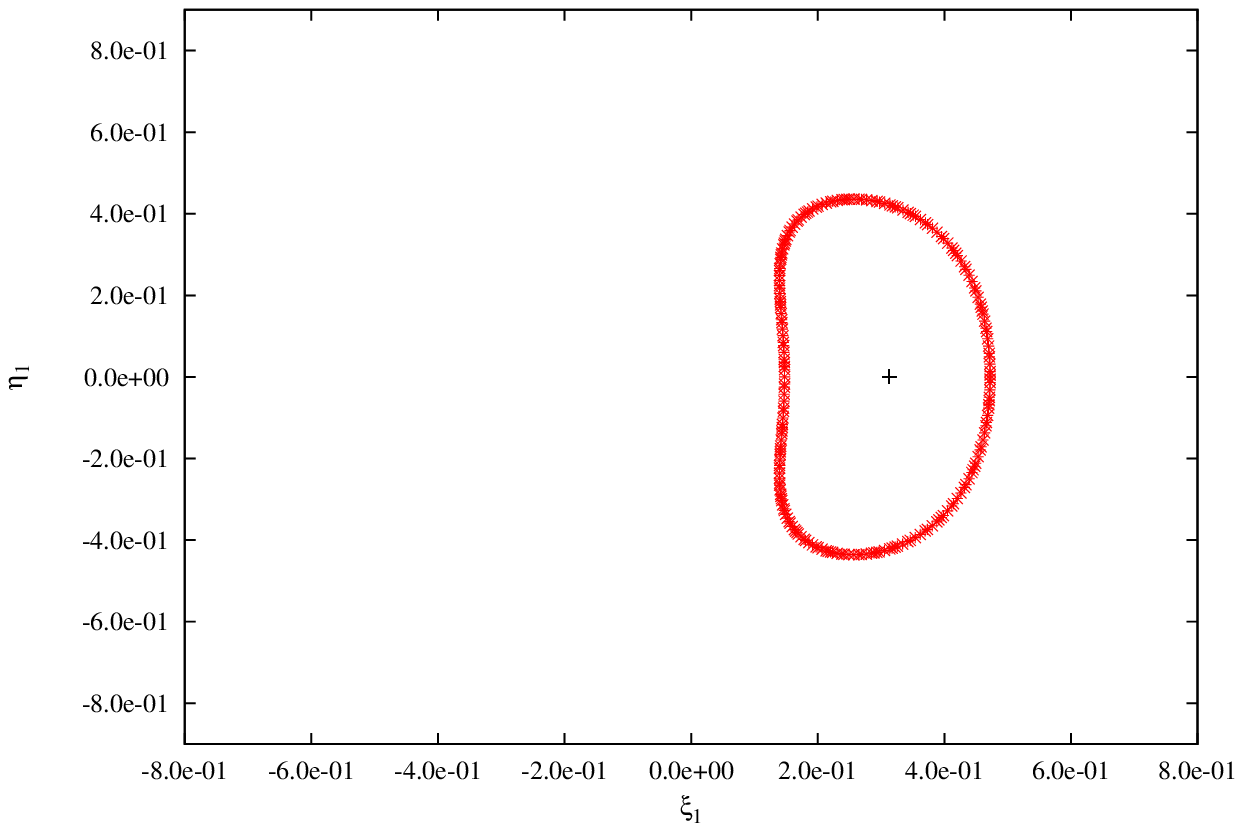}
      \caption{On the left, Poincar\'e sections that are corresponding
        to the hyperplane $\eta_2=0$ (with the additional condition
        $\xi_2>0$) and are generated by the flow of the Hamiltonian
        secular model $H^{({\rm sec})}$, given in~\eqref{frm:hsec} at
        order two in the masses for the exoplanetary system
        $\upsilon$~Andromed{\ae}; the orbit in red refers to the
        motion starting from initial conditions of the same type of
        those considered in
        Figs.~\ref{maxecc_maxdeltaomega}--\ref{orbU} with the
        additional choice of the initial mean anomalies, that have
        been fixed so that $M_1(0)=M_2(0)=0^\circ$. On the right
        panel, the same orbit in red is shown: approximately at its
        centre the symbol $+$ represents the orbit of a
        one-dimensional elliptic torus (that reduces to a fixed point
        in these Poincar\'e sections). See the text for more details.}
      \label{sez_Poin_sec}
  \end{figure}

A few dynamical features of the Hamiltonian model defined by $H^{({\rm
    sec})}$ are summarised in the plots reported in
Fig.~\ref{sez_Poin_sec}. The orbit plotted in red in both panels
refers to a set of initial conditions of the same type with respect to
those considered in the previous
Section~\ref{sec:Ups-Andromedae_overview}. In detail, the initial
values of the mean anomalies have been set so that
$M_1(0)=M_2(0)=0^\circ$, while the other initial conditions are taken
from the mid values of Table~\ref{tab:orbel-with-errors} (for what
concerns $\iota(0)$, $\omega(0)$ and $\Omega(0)$ only) and from
Table~\ref{tab:orbel-partial} (for the remaining data).  Since the
Poincar\'e sections are plotted in correspondence to the hyperplane
$\eta_2=0$ (with the additional condition $\xi_2>0$) and the canonical
variables $(\vet \xi,\vet \eta)$ appearing in formula~\eqref{frm:hsec}
are close to those defined in~\eqref{frm:poin-var}, then we can assume
that on the surface of section $\omega_2\simeq 0$. In the left panel
of Fig.~\ref{sez_Poin_sec}, therefore, the difference of the
pericentre arguments $\omega_2-\omega_1$ is evaluated by the polar
angle, whose width is measured, as usual, with respect to the set of
the positive abscissas, i.e., $\{(\xi_1>0,\eta_1=0)\}$.  Thus, we can
easily appreciate that this angle is librating around $0^{\circ}$ also
in the case of the secular model, in agreement with the corresponding
plots reported in Figs.~\ref{maxecc_maxdeltaomega}--\ref{orbU}, that
refer to the dynamics of the complete planetary system. By taking into
account the fact that the nodes are opposite in the Laplace frame,
this means that the pericentres of $\upsilon$~And~\emph{c} and
$\upsilon$~And~\emph{d} are in the so-called apsidal locking regime in
the vicinity of the anti-alignment of the pericentres.  It is easy to
remark that the Poincar\'e sections plotted in red (that are
corresponding to the motion starting from the initial conditions we
have chosen to consider) are orbiting around a fixed point, whose
presence is also highlighted in the right panel of
Fig.~\ref{sez_Poin_sec}. Let us recall that all the Poincar\'e
sections reported in Fig.~\ref{sez_Poin_sec} refer to the same level
of energy, say $E$, corresponding to the set of the initial conditions
we have previously described. Since $H^{({\rm sec})}$ is a two degrees
of freedom Hamiltonian, the manifold labelled by such a value of the
energy will be three-dimensional; in other words, by plotting the
Poincar\'e sections, we automatically reduce by one the dimensions of
the orbits. This is the reason why a fixed point actually corresponds
to a periodic orbit. Since such a fixed point with positive value of
the abscissa is surrounded by closed curves, then we can argue that
such a periodic orbit is linearly stable for what concerns the
transverse dynamics. This means that it can be seen as a
one-dimensional elliptic torus. Therefore, we can conclude that the
orbit which intersects the hyperplane $\eta_2=0$ in correspondence
with the red dots is actually winding around a linearly stable
periodic orbit, by remaining in its vicinity.  This explains why it
can be convenient to adopt a strategy based on two different
algorithms: the first one refers to the elliptic torus (that
corresponds to a fixed point in the Poincar\'e sections) and provides
a good enough approximation to start the second computational
procedure that constructs the final KAM torus (which shall include
also the points marked in red in Fig.~\ref{sez_Poin_sec}).

\subsection{Construction of the normal form for a 1D elliptic torus}\label{sbs:ell_torus}
It is now convenient to introduce a new set of canonical coordinates
by including among them also an angle which describes the libration of
the difference of the pericentre arguments, i.e.,
$\omega_2-\omega_1\,$.  For such a purpose, we first introduce a set
of action-angle variables $(\vet\Jscr,\vet\psi)$ via the canonical
transformation
\begin{equation}
\xi_j=\sqrt{2\Jscr_j}\cos \psi_j\ , \qquad \eta_j=\sqrt{2\Jscr_j}\sin \psi_j\ ,
\qquad \forall\ j=1,2,
\label{eq:azang}
\end{equation}
being $(\vet\xi,\vet\eta)$ the variables appearing as arguments of the
secular Hamiltonian $H^{({\rm sec})}$ defined in~\eqref{frm:hsec}.
Then, we define a new set of variables $(\vet I, \vet\theta)$
such that
\begin{equation}
\theta_1 = \psi_1 -\psi_2\ , \quad \theta_2 = \psi_2\ , \quad
I_1 = \Jscr_1\ ,  \quad I_2 = \Jscr_2 + \Jscr_1\ .
\label{eq:def-azang-Iphi}
\end{equation}
In view of the discussion included in the previous subsection, we have
that the angle $\theta_1\simeq\omega_2-\omega_1$ is expected to
librate in the model under consideration.  We now move to (new)
canonical polynomial variables $(\vet x, \vet y)$ defined as
\begin{equation}
x_j = \sqrt{2 I_j}\cos \theta_j\ , \qquad
y_j=\sqrt{2 I_j}\sin \theta_j\ , \qquad \forall\ j=1,2\ .
\label{eq:def-dalembcoord-xy}
\end{equation}
Let us also remark that making Poincar\'e sections with respect to the
hyperplane $\eta_2=0$, when $\xi_2>0$ is equivalent to impose
$\psi_2=0$, because of the definitions in~\eqref{eq:azang}. Therefore,
looking at
formul{\ae}~\eqref{eq:def-azang-Iphi}--\eqref{eq:def-dalembcoord-xy},
one can easily realise that the drawing in the left panel of
Fig.~\ref{sez_Poin_sec} can be seen as a plot of the
Poincar\'e sections in coordinates $(x_1\,,\,y_1)$ with respect to
$y_2=0$ and with the additional condition $x_2>0$. 
By a simple numerical method, we can
easily determine the initial condition $( \vet x^\star, \vet y^\star)$
that is in correspondence with a Poincar\'e section and generates a
periodic solution. We can now subdivide the variables in two different
pairs. The first one is given by $(p,q)\in \reali\times \toro$,
i.e., the action-angle pair describing the periodic motion.  Thus,
we rename the angle $\phi_2$ as $q$, while the action is obtained by
translating the origin of $I_2$ so that $p= I_2 -
\big((x_2^{\star})^2+(y_2^{\star})^2\big)/2$. For what concerns the
second pair of canonical coordinates, we start from the polynomial
variables $(x_1,y_1)$ in order to describe the motion transverse to
the periodic orbit. It is now convenient to rescale the
transverse variables $(\bar x_1,y_1)$, being $\bar x_1= x_1 -
x_1^\star$, in such a way that the Hamiltonian part which is quadratic
in the new variables $(x,y)$ and does not depend on $(p,q)$ is in the
form $\Omega^{(0)}(x^2 + y^2)/2$. This rescaling can be done by a
canonical transformation as the quadratic part does not have any mixed
term $\bar x_1 y_1$ and the coefficients of $\bar x_1^2$ and $y_1^2$
have the same sign, because of the proximity to an elliptic
equilibrium point. Thus, since such a quadratic part is in the
preliminary form $a\bar x_1^2+ b y_1^2$, it suffices to define the new
variables $(x,y)$ as $x = \sqrt[4]{\frac a b}\, \bar x_1, \ y =
\sqrt[4]{\frac b a}\, y_1\,$. Finally, we introduce the second pair of
canonical coordinates $(J,\phi)\in \reali_+\cup\{0\}\times \toro$ so
that $x=\sqrt{2 J}\cos\phi$ and $y=\sqrt{2 J}\sin\phi$.

After having performed all the canonical transformation described
above, the Hamiltonian can be written in the following way:
\begin{equation}
\label{frm:Hscr^(0)}
\vcenter{\openup1\jot 
\halign{
$\displaystyle\hfil#$&$\displaystyle{}#\hfil$&$\displaystyle#\hfil$\cr
  \Hscr^{(0)}(p, q, J, \phi) =
  & \Escr^{(0)}+\nu^{(0)} p+\Omega^{(0)} J+h(p, J,\phi)
  + \epsilon f^{(0)}(p, q, J, \phi)
  \ ,
  \cr
}}
\end{equation}
where $\Escr^{(0)}$ is constant (that is close to the energy value of
the wanted periodic orbit), $\nu^{(0)}$ and $\Omega^{(0)}$ are angular
velocities, the function $h(p, J,\phi)=\Oscr(\|(p,J)\|^{3/2})$ when
the action vector\footnote{Because of the change of coordinates which
  introduces the canonical pair of variables $(J,\phi)$, i.e.,
  $x=\sqrt{2 J}\cos\phi$ and $y=\sqrt{2 J}\sin\phi$, also semi-integer
  powers of $J$ can appear in the expansion~\eqref{frm:Hscr^(0)} of
  the Hamiltonian $\Hscr^{(0)}$.} $(p,J)\to \vet 0$ and $\epsilon
f^{(0)}(p, q, J, \phi)$ is a generic perturbing term, with $\epsilon$
playing the role of the small parameter.  If such a perturbation is
small enough, then it is possible to successfully perform a
normalization algorithm, which allows to construct another canonical
transformation $\Phi$ that conjugates the initial Hamiltonian
$\Hscr^{(0)}$ to $\Hscr^{(\infty)}=\Hscr^{(0)}\circ\Phi$ having the
following (normal) form:
\begin{equation}
\label{frm:Hscr^(infty)}
\vcenter{\openup1\jot 
\halign{
$\displaystyle\hfil#$&$\displaystyle{}#\hfil$&$\displaystyle#\hfil$\cr
  \Hscr^{(\infty)}(P, Q, X, Y) =
  & \Escr^{(\infty)}+\nu^{(\infty)}P+
       \frac{\Omega^{(\infty)}}{2}\big(X^2+Y^2\big)+
       \Rscr(P, Q, X, Y)
  \ ,
  \cr
}}
\end{equation}
where $\Escr^{(\infty)}$ is constant, $\nu^{(\infty)}$ and
$\Omega^{(\infty)}$ are angular velocities and the remainder $\Rscr$
is such that $\Rscr(   P,   Q,   X,   Y)=o\big(|P|+
\|(  X,  Y)\|^2\big)$, when $(   P,   X,   Y)\to(  
0,   0,   0)$. Therefore, one can easily check that
\begin{equation}
  \label{frm:soluzione-su-toro-ellittico}
  (    P(t),     Q(t),    X(t),    Y(t)) =
  \big(    0,     Q(0) +     \nu^{(\infty)} t,     0,    0 \big)
\end{equation}
is a solution of the Hamilton equations, since the function
$\Hscr^{(\infty)}$, contains terms of type $\Oscr(P^2)$, $\Oscr(|P|\|(
X, Y)\|)$ and $\Oscr(\|( X, Y)\|^3)$ only, except for its main part
(that is made by a constant, a linear term in $P$ and another
quadratic in both $X$ and $Y$).  Because of this remark, it is evident
that the 1D manifold $\big\{( P, Q, X, Y)\,:\ P = 0,\> Q \in\toro,\> X
= Y = 0 \big\}$ is invariant. The energy level of such a solution is
equal to $\Escr^{(\infty)}$.  The elliptical character is given by the
fact that, in the remaining degree of freedom, the transverse dynamics
is given by an oscillatory motion whose period tend to the value
$2\pi/\Omega^{(\infty)}\,$, in the limit of $( P, X, Y)\to( 0, 0, 0)$.
Of course, this is due to the occurrence of the term
$\Omega^{(\infty)}(X^2 + Y^2)/2$ which overwhelms the effect of the
remainder $\Rscr$ in the so-called limit of small oscillations.

In the case under study, dealing with the exoplanetary system
$\upsilon$~Andromed{\ae}, the normalization algorithm can be adapted
so as to construct the 1D elliptic torus with a value of the parameter
$\Escr^{(\infty)}$ equal to the energy level of the Poincar\'e
sections (see~\cite{Car-Loc-San-Vol-2022}). In the right panel of
Fig.~\ref{sez_Poin_sec} all the intersections of the corresponding
orbit with the Poincar\'e surface $\eta_2=0$ are marked with a black
cross. Of course, they perfectly superpose each other in a single
fixed point corresponding to the wanted periodic
orbit. In~\cite{Caracciolo-2021}, the normalization algorithm we have
adopted to construct elliptic tori is fully described and its
convergence is thoroughly analysed from a theoretical point of view.
In short, such a procedure can be made in strict analogy with the
construction of the Kolmogorov normal form. In fact, it can be
formulated in such a way to introduce a sequence of Hamiltonians
$H^{(r)}$ that are iteratively defined by a normalization step that is
mainly composed by three Lie series: the first aims to reduce the
perturbation that is not depending on the actions $(p,J)$; the second
achieves the same with the terms proportional to $\sqrt{J}$; also the
third has the same goal for what concerns the terms that are linear in
$p$ or in $J$. We remark that in the normal form Hamiltonian
$\Hscr^{(\infty)}$ written in~\eqref{frm:Hscr^(infty)} we have
expressed the dynamics that is transverse to the 1D elliptic torus in
terms of the normalised canonical coordinates $(X,Y)$ of polynomial
type (instead of using action--angle variables), in order to highlight
the existence of the periodic
solution~\eqref{frm:soluzione-su-toro-ellittico}.

  \begin{figure}
    \centering
      \includegraphics[clip, height=4.6cm,]{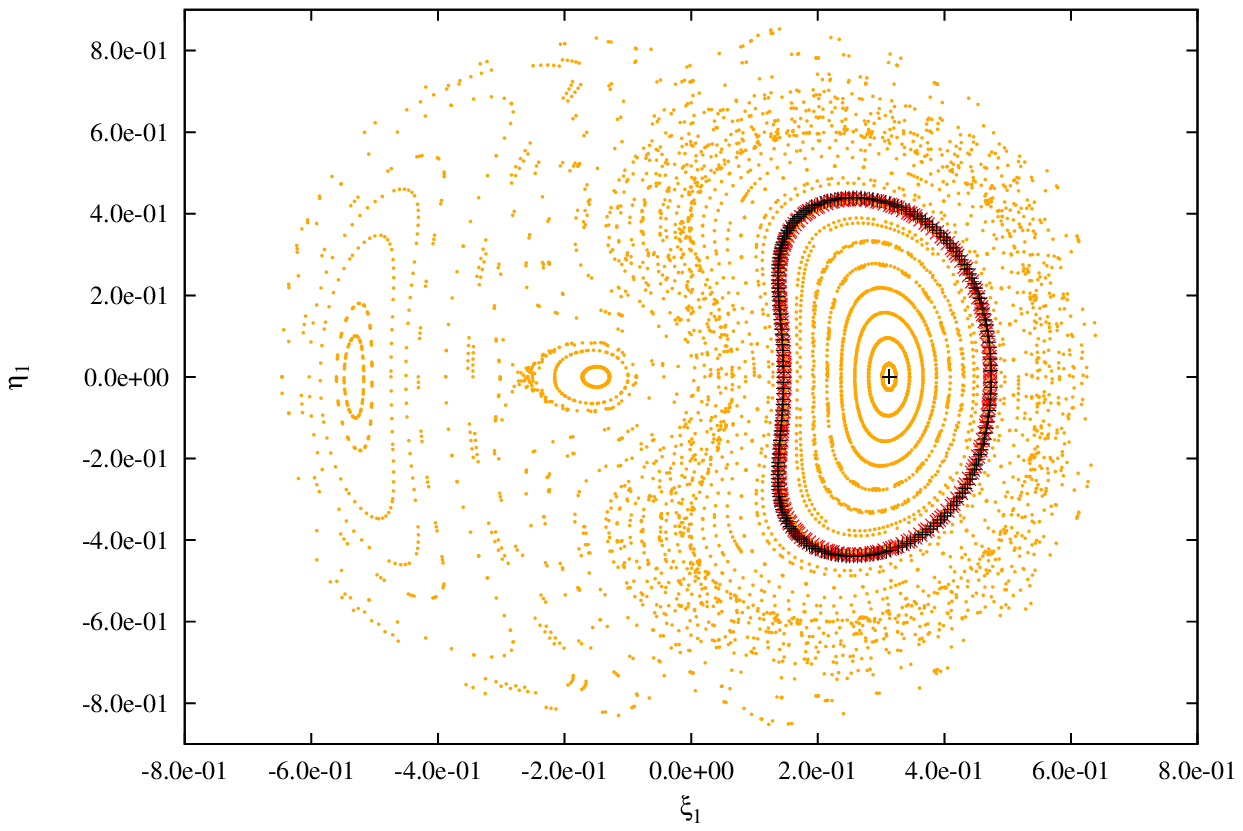}
      \includegraphics[clip, height=4.6cm,]{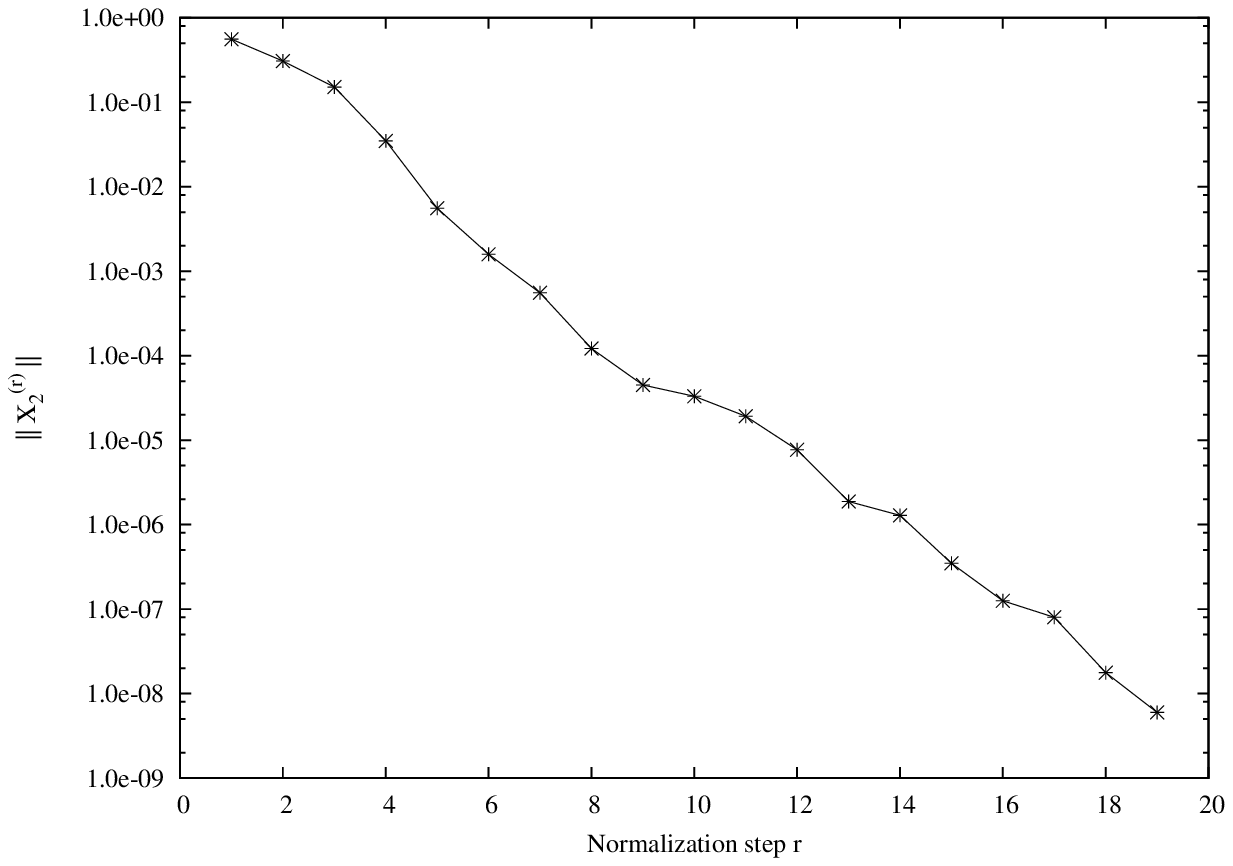}
      \caption{On the left, comparisons between the Poincar\'e
        sections generated by two different initial conditions. The
        first ones are marked in red and are exactly the same as those
        appearing in the left panel of Fig.~\ref{sez_Poin_sec} (where
        they are plotted in red as well). The second ones are marked
        in black and correspond to the orbit on the invariant KAM
        torus. The other ``background'' Poincar\'e sections are defined
        in the same way as those reported in Fig.~\ref{sez_Poin_sec};
        in particular, the dots plotted in blue there are located
        exactly in the same positions as those marked in orange
        here. The black symbol~$+$ refers to the orbit of the 1D
        elliptic torus also here.  On the right, the behaviour of
        $\|\chi_2^{(r)}\|$ is plotted as a function of the
        normalization step $r$.}
      \label{sez_Poin_completo_sec+normegenKAM}
  \end{figure}

\subsection{Final construction of the invariant KAM torus}\label{sbs:KAM_torus}
It is now convenient to express also the second pair of canonical
coordinates appearing in the normalised
Hamiltonian~\eqref{frm:Hscr^(infty)} in the form of action--angle
variables, i.e., we introduce $(I,\Theta)$ so that
$X=\sqrt{2I}\cos\Theta$ and $Y=\sqrt{2I}\sin\Theta$.
A very simple canonical change of variables, i.e.,
\begin{equation}
  \label{frm:final-trasl}
  p_1 = P \>, \quad 
  q_1 = Q \>, \quad
  p_2 = I - I^\star \>, \quad 
  q_2 = Q\>,
\end{equation}
is now enough in order to transform the Hamiltonian $\Hscr^{(\infty)}$
(introduced at the end of the previous subsection) to
$H^{(0)}(\vet{p},\vet{q})=\vet{\nu}\cdot\vet{p}+
h^{(0)}(\vet{p},\vet{q})+\epsilon f^{(0)}(\vet{p},\vet{q})$, that is
in a suitable form to start the classical normalization algorithm that
is the base of the proof scheme of KAM theorem. In a first
approximation, the translation constant can be determined as
$I^\star=(X_0^2+Y_0^2)/2$, where $(X_0\,,\,Y_0)$ are values of the
canonical coordinates $(X,Y)$ corresponding to the initial
conditions. Moreover, as a preliminary step we determine the angular
velocity vector $\vet{\nu}$ by using the frequency analysis
method. The choice of $I^\star$ can be optimised by applying a Newton
method, so as to approach as much as possible the vector $\vet{\nu}$
(see~\cite{Car-Loc-San-Vol-2022}).
Fig.~\ref{sez_Poin_completo_sec+normegenKAM} highlights that the
algorithm constructing the Kolmogorov normal
form~\eqref{eq:Kolmogorov-normal-form} is successful also for the
initial conditions considered in the present section, i.e., with mean
anomalies fixed so that $M_1(0)=M_2(0)=0^\circ$, while the other
initial conditions are taken from the mid values of
Table~\ref{tab:orbel-with-errors} (for what concerns $\iota(0)$,
$\omega(0)$ and $\Omega(0)$ only) and from
Table~\ref{tab:orbel-partial} (for the remaining data).  In the left
panel the Poincar\'e sections that are plotted (in red) during the
numerical integration of the equations of motion related to the
Hamiltonian~\eqref{frm:hsec} perfectly superpose to the orbit produced
by composing all the canonical transformations briefly described in
the present section, which is marked in black. The right panel of
Fig.~\ref{sez_Poin_completo_sec+normegenKAM} clearly shows the
regularity of the decrease of the norms of the generating functions
(which are computed by simply adding up the absolute values of all the
Taylor--Fourier coefficients). This gives a clear numerical indication
of the convergence of the computational procedure in the case under
study dealing with the exoplanetary system $\upsilon$~Andromed{\ae}.

We stress that the importance of the translation constant $I^\star$ is
crucial. Indeed, the abundance of the KAM manifolds surrounding an
invariant torus is an increasing function of the inverse of the
distance from said torus (as it has been shown, e.g.,
in~\cite{Mor-Gio-1995}). This is in agreement with the rather well
known fact that the small parameter $\epsilon$, which enters in the
definition of the Hamiltonian $H^{(0)}$, is proportional to the shift
value $I^\star$ (see, e.g., \cite{Gio-Loc-San-2017}). Therefore, also
the rate of the exponential decrease of the generating functions
depends on $I^\star$: the smaller the latter the faster the former. In
other words, we can also say that the invariant tori surrounding a
reference one are more and more {\it robust} when the shift value
$I^\star$ tends to zero. This means that larger and larger additional
perturbing terms are needed in order to destroy this invariant
structure for $I^\star\to 0$. Of course, all these remarks still hold
true also when the reference torus (corresponding to $I^\star=0$) is
of elliptic type, as it is in the case of the periodic orbit
$(P(t),Q(t),X(t),Y(t)) = \big(0,Q(0)+\nu^{(\infty)}t,0,0\big)$ that is
obviously invariant with respect to the Hamiltonian flow of the normal
form $\Hscr^{(\infty)}$ written in~\eqref{frm:Hscr^(infty)}.

\section{The criterion of the minimal area as a robustness indicator}
\label{sec:num-criterion}

\subsection{Motivation and definition}\label{sbs:def-crit}
In the final discussion at the end of the previous section, we have
explained why the shift value $I^\star$ appearing in the canonical
transformation~\eqref{frm:final-trasl} can be considered as a good
indicator of the dynamical robustness of an eventually existing KAM
torus. However, such a concept is not easy to use in the context of
numerical explorations, because its computation would require to
preliminarily construct the normal forms we have previously described.
Here, we are going to make the effort to reformulate our approach in a
way that is far more handy in view of practical applications.

Firstly, let us remark that from the
definition~\eqref{frm:final-trasl} it immediately follows that the
shift value $I^\star$ has the physical dimensions of an action.  Let
us also recall that in Hamiltonian systems having one degree of
freedom, the action is usually introduced as the area contoured by a
closed orbit (see, e.g., \S~50 of~\cite{Arnold-book}). Since the
action $I=(X^2+Y^2)/2$ is a sort of squared distance in the pair of
canonical coordinates $(X,Y)$ which describe the transverse dynamics
with respect to the 1D elliptic torus, then it looks rather natural to
transfer the role of robustness indicator from the quantity $I^\star$
to the area enclosed by an orbit in the Poincar\'e sections. Let us
directly refer to Fig.~\ref{sez_Poin_sec} in order to fix the
ideas. We recall that we have adopted the non-normalised canonical
coordinates $(\vet\xi,\vet\eta)$ to plot those Poincar\'e sections,
instead of $(X,Y)$ that are much more expensive to
compute. Nevertheless, in the hyperplane $\eta_2=0$ (after having
fixed the energy level) the pair $(\xi_1\,,\,\eta_1)$ evidently
describes a manifold that is transverse to the 1D elliptic torus,
which is located by a fixed point marked with a black cross in the
right panel. Since all the invariant tori winding around that periodic
orbit describe Poincar\'e sections which are enclosing each other,
then we can assume that the area embraced by the Poincar\'e sections
is proportional to the distance (in action) from the elliptic
torus. Therefore, by combining all the arguments explained at the end
of the previous section with those discussed at the beginning of the
present one, it is natural to assume that \emph{an invariant torus is
  as more robust as smaller is the area contoured by the corresponding
  Poincar\'e sections}.

\noindent We now come to the approximated evaluation of such an area.
By focusing our attention on the Poincar\'e sections marked in red in
both panels of Fig.~\ref{sez_Poin_sec}, we can say that the
corresponding area is nearly equal to
\begin{equation}
  \label{frm:to-def-area}
  \left(\max_t\big\{\xi_1(t)\big\} - \min_t\big\{\xi_1(t)\big\}\right)
  \left(\max_t\big\{\eta_1(t)\big\} - \min_t\big\{\eta_1(t)\big\}\right)\ .
\end{equation}
Let us recall that $\xi_1$ and $\eta_1$ are proportional to
$e_1\cos\omega_1$ and $e_1\sin\omega_1\,$, respectively, as determined by
the definitions~\eqref{frm:poin-var}. Therefore, we can assume that
also the area written in the formula above is proportional to
\begin{equation}
  \label{def:area-Poinc-sect}
  \Ascr =
    \Big[\big(e_{1; {\rm max}}\big)^2 - \big(e_{1; {\rm min}}\big)^2\Big]
    \max_t \big|\omega_1(t)-\omega_2(t)\big|\ ,
\end{equation}
where the meaning of the new symbols we have just introduced is $e_{1;
  {\rm max}}=\max_t\big\{e_1(t)\big\}$ and $e_{1; {\rm
    min}}=\min_t\big\{e_1(t)\big\}$.  Moreover, in order to write the
definition of the quantity $\Ascr$ above as an approximation of the
action surface written in formula~\eqref{frm:to-def-area}, we have
also assumed that (by symmetry reasons) both the extremals
$\max_t\{\xi_1(t)\}$ and $\min_t\{\xi_1(t)\}$ are in correspondence
with $\omega_1=0$, while we have evaluated the width
$\max_t\{\eta_1(t)\} - \min_t\{\eta_1(t)\}$ with a circular arc
centred in the origin $(\xi_1\,,\,\eta_1)=\vet{0}$ of the frame of the
Poincar\'e surface. We remark that the half-width of that arc is
evaluated by referring to $\big|\omega_1(t)-\omega_2(t)\big|$, because
$\omega_2$ is equal to zero in the region of the Poincar\'e surface
with $\xi_1>0$ and we want to evaluate the quantity $\Ascr$ for any
motion in librational regime with respect to the difference of the
pericentre arguments.

We can summarise all the discussion above by formulating the following

\begin{description}
  \item
    \emph{robustness criterion: we assume that a quasi-periodic
      Hamiltonian motion describing an invariant torus is as more
      robust as smaller is the corresponding quantity $\Ascr$ defined
      in~\eqref{def:area-Poinc-sect}}.
\end{description}

It is quite evident that the statement above requires to minimise the
area enclosed by the Poincar\'e sections, when we look for the most
robust orbit originating from a set of possible initial conditions.
For short, hereafter, we will refer to that as the criterion of the
``minimal area''.

\subsection{An application to the $\upsilon$~Andromed{\ae} extrasolar planetary system}\label{sbs:appl-crit}
In spite of the fact that we have constantly referred to a
\emph{secular} model in order to introduce and motivate our robustness
criterion, we emphasise that its formulation is so flexible that it
can be applied also to the study of the \emph{complete} planetary
dynamics of extrasolar systems. As we have claimed since the
Introduction of the present work, we are going to select a set of
initial conditions that is corresponding to an orbital configuration
of the $\upsilon$~Andromed{\ae} three-body model which is extremely
stable.  In our opinion, looking for \emph{robust} invariant tori with
a numerical criterion inspired by the secular dynamics has a twofold
meaning. Firstly, they have more chances to persist when the
perturbing effects due to the fast dynamics are taken into account; as
we have shown in Section~\ref{sec:Ups-Andromedae_overview}, chaotic
motions compatible with the initial conditions are not rare in a
probabilistic sense. Moreover, \emph{robust} invariant tori describing
the orbits $\upsilon$~And~\emph{c} and $\upsilon$~And~\emph{d} are
expected to stay within a dynamically stable region of the phase space
also when the effects due to $\upsilon$~And~\emph{b} and/or
$\upsilon$~And~\emph{e} are included in the model.

\begin{figure}
  \vspace*{-0.6 cm}
  \begin{center}
  \parbox[c]{6.4cm}{\includegraphics[height=4.6cm,trim= 0cm 0.2cm 0cm 0cm]{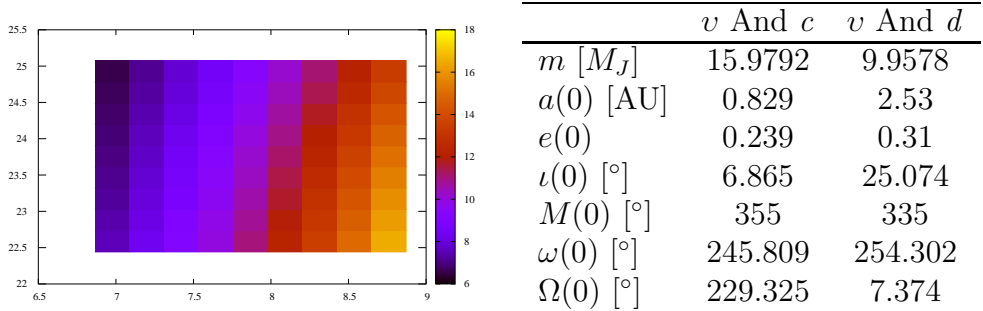}}
  \hspace{0.5cm}
    \begin{tabular}{l c c}
    \hline
    & $\upsilon$~And~\emph{c} & $\upsilon$~And~\emph{d} \\ 
    \hline
    $m$ [$M_J$] & $15.9792$ & $ 9.9578 $ \\
    $a(0)$ [AU] & $0.829$ & $2.53$ \\
    $e(0)$ & $0.239$ & $0.31$\\
    $\iota(0)$ [$^\circ$]& $6.865$ & $25.074$ \\ 
    $M(0)$ [$^\circ$] & $355$ & $335$ \\
    $\omega(0)$ [$^\circ$] & $245.809$ & $254.302$ \\
    $\Omega(0)$ [$^\circ$] & $229.325$ & $7.374$\\
    \hline
  \end{tabular}
  \end{center} 
  \centering
  \centering
  \caption{On the left, colour-code plot of the area $\Ascr$ for
    different initial values of the inclinations $\iota_1(0)$ and
    $\iota_2(0)$. See formula~\eqref{def:area-Poinc-sect} for the
    definition of the quantity $\Ascr$. On the right, Table including
    the values of the masses and the initial conditions
    expected to correspond to the most robust planetary orbit
    compatible with the observed data available for
    $\upsilon$~And~\emph{c} and $\upsilon$~And~\emph{d}, according to
    the criterion of the minimal area. }
  \label{fig:crit_min_area}
\end{figure}

In order to avoid the extensive study of a grid of initial conditions
having a too high dimensionality, we will split our analysis in three
different layers. As a first step, we consider initial conditions such
that the mean anomalies are fixed so that $M_1(0)=M_2(0)=0^\circ$,
while ${\vet\omega}(0)$ and ${\vet\Omega}(0)$ are taken from the
corresponding mid values of Table~\ref{tab:orbel-with-errors};
moreover, the assumed values of ${\vet a}(0)$, ${\vet e}(0)$ and
minimal masses come from Table~\ref{tab:orbel-partial}; the initial
data are completed by covering the range of values of $\iota_1(0)$ and
$\iota_2(0)$ which is reported in Table~\ref{tab:orbel-with-errors}
with a regular grid of 10x10 points. For each of these~$100$ initial
conditions, we numerically integrate the equations of motion, by using
the symplectic method $\Sscr\Bscr\Ascr\Bscr_{\Cscr 3}$ (also here we
adopt the same integrator as in
Section~\ref{sec:Ups-Andromedae_overview}, which is described
in~\cite{Las-Rob-01}, with the same total timespan and integration
step, that are $10^5$~yr and $0.02$~yr, respectively), and we compute
the corresponding value of the numerical indicator $\Ascr$. The
results are reported in the left panel of
Fig.~\ref{fig:crit_min_area}. A straightforward application of the
minimal area criterion allows us to conclude that the initial
conditions that are expected to correspond to the most robust
planetary orbit are such that
\begin{equation}
  \label{frm:incl+mass}
  \iota_1(0)=6.865^\circ\ ,
  \ \ \iota_2(0)=25.074^\circ\ ,
  \quad \Longrightarrow \quad
  m_1 =15.9792\,M_J\ ,
  \ \ m_2 = 9.9578\,M_J\ .
\end{equation}
The left panel of Fig.~\ref{fig:crit_min_area} clearly shows a rather
surprising result: the most robust configurations correspond to the
minimal value of the initial inclination $\iota_1(0)$ and, thus, to
the maximal value of the $\upsilon$~And~\emph{c} mass (i.e., $\simeq
16\,M_J$).  This conclusion is in agreement with a similar analysis
that has been performed in~\cite{Car-Loc-San-Vol-2022}, by studying
the ratio between the norm of the last- and first-computed generating
function, among those reported in a graph analogous to that appearing
in the right panel of
Fig.~\ref{sez_Poin_completo_sec+normegenKAM}. The decrease rate of the
sequence of the generating functions $\{\chi_2^{(r)}\}_{r\ge 1}$
(which are defined by the normalization algorithm eventually leading
to the final Kolmogorov normal form) has been been firstly adopted as
a robustness indicator starting from~\cite{Vol-Loc-San-2018}. We
stress that this our new result looks to be rather unexpected when
compared with the existing ones in the scientific literature: none of
the four stable (and prograde) orbital configurations reported in
Table~3 of~\cite{Deitrick-et-al-2015} is such that the
$\upsilon$~And~\emph{c} mass is greater than $11\,M_J$, that is below
the lowest possible value of $m_1=1.91/\sin(\iota_1(0))\>M_J$, where
$1.91\,M_J$ is the minimal mass of $\upsilon$~And~\emph{c} taken from
Table~\ref{tab:orbel-partial} and its initial inclination $\iota_1(0)$
is ranging in the corresponding interval reported in
Table~\ref{tab:orbel-with-errors}.

We continue our analysis by studying a second layer.  We now consider
initial conditions such that the mean anomalies are still fixed so
that $M_1(0)=M_2(0)=0^\circ$, while ${\vet a}(0)$, ${\vet e}(0)$ are
taken from Table~\ref{tab:orbel-partial} and the values of the initial
inclinations and masses are as written in
formula~\eqref{frm:incl+mass}. In this second layer of analysis, the
initial data are completed by covering the range of values of the
angles ${\vet\omega}(0)$ and ${\vet\Omega}(0)$ with a regular 4D~grid.
Since the uncertainties on the knowledge of both the pericentre
arguments and the longitudes of the node are not so large, we limit
ourselves to define a grid which considers for each of the angles
$\omega_1(0)$, $\omega_2(0)$, $\Omega_1(0)$, and $\Omega_2(0)$ just
three possible values that are the minimum, the mid-point and the
maximum of the corresponding values range reported in
Table~\ref{tab:orbel-with-errors}, respectively.
For each of the so defined~$81$ initial conditions, we perform the
same type of numerical integration we have described above.  In this
case, the application of the minimal area criterion leads to the
conclusion that the initial values of ${\vet\omega}$ and
${\vet\Omega}$ that are expected to correspond to the most robust
orbit are those reported in the Table included on the right of
Fig.~\ref{fig:crit_min_area}.

\begin{figure}
  \vspace*{-0.4 cm}
  \begin{center}
  \includegraphics[clip, height=4.6cm,]{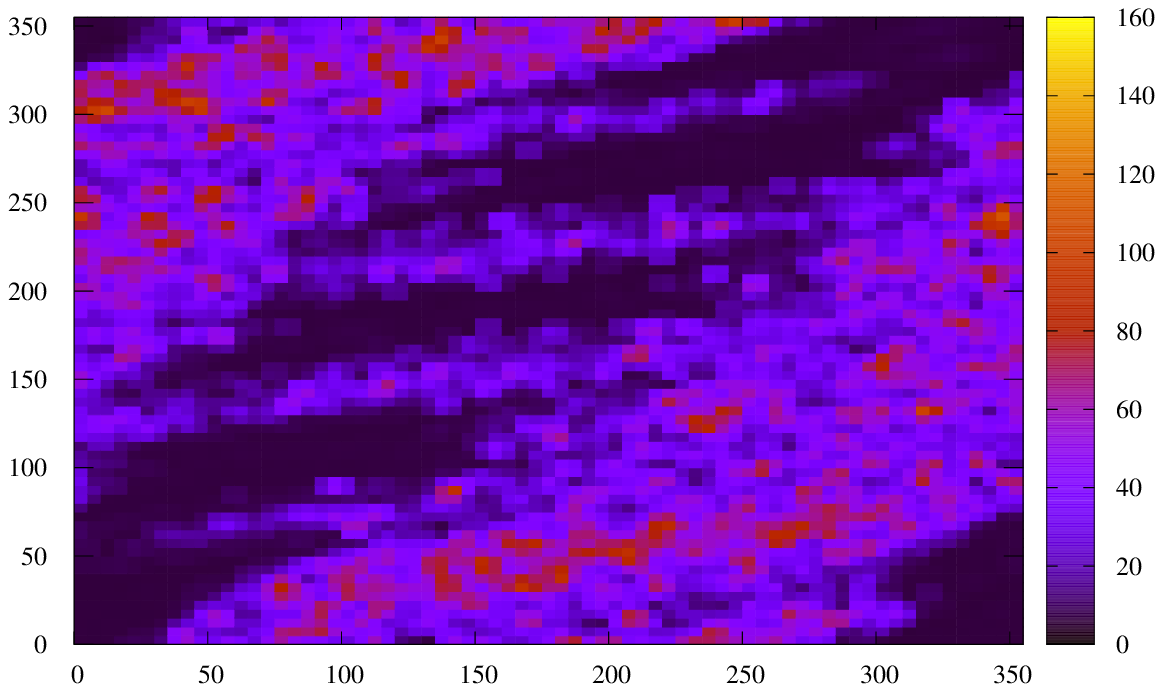}
  \includegraphics[width=5.2cm,trim= 0cm -1.2cm 0cm 0cm]{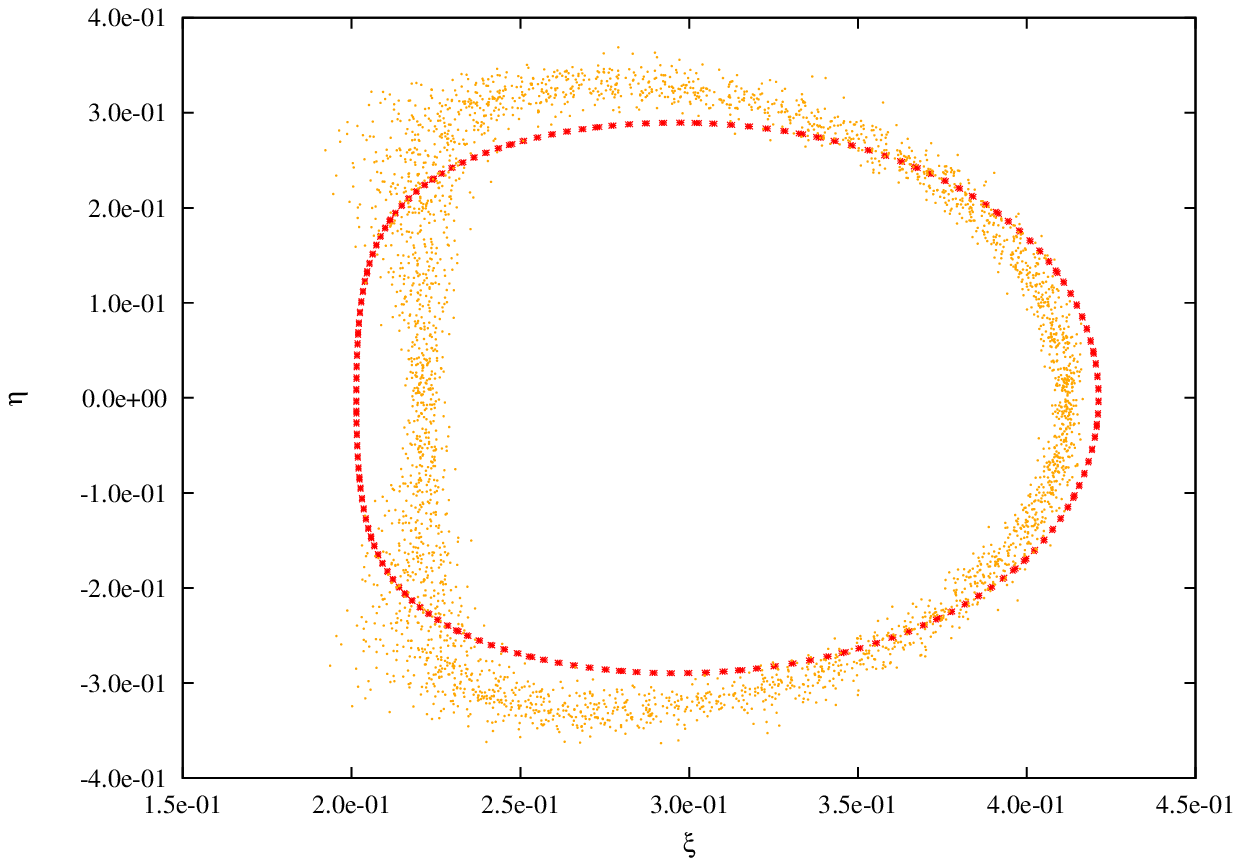}
  \end{center}
      \caption{On the left, colour-code plot of the area $\Ascr$ for
        different initial values of the mean anomalies $M_1(0)$ and
        $M_2(0)$. See formula~\eqref{def:area-Poinc-sect} for the
        definition of the quantity $\Ascr$. On the right, in orange,
        plot of the Poincar\'e sections that are corresponding to the
        hyperplane $\eta_2=0$ (with the additional condition
        $\xi_2>0$) and are generated by the flow of the complete
        Hamiltonian model of the exoplanetary system
        $\upsilon$~Andromed{\ae}; such a motion is started from the
        initial conditions listed in the Table included on the
        right of Fig.~\ref{fig:crit_min_area}. In red, plot of the
        Poincar\'e sections generated by the flow of the secular model
        $H^{({\rm sec})}$, given in~\eqref{frm:hsec} and starting from
        initial conditions generated by those same values of the
        orbital elements.}
   \label{fig:confr_sec+compl}
\end{figure}

We come now to the description of the third layer of our analysis.  In
this last case, we consider the values of the planetary masses
$m_1\,$, $m_2$ and the initial conditions for the orbital elements
$\vet{a}$, $\vet{e}$, $\vet{\iota}$, $\vet{\omega}$ and $\vet{\Omega}$
as they are given in the Table included on the right of
Fig.~\ref{fig:crit_min_area}, while we make the coverage of all the
possible initial values of the mean anomalies
$\big\{\big(M_1(0),M_2(0)\big)\big\}\in
[0^\circ,360^\circ]\times[0^\circ,360^\circ]$ by means of a regular 2D
grid with a grid-step of $5^\circ$. Once again, for each of these
$72^2=5184$ different initial conditions, we perform the same type of
numerical integration we have described above.  For all of them, we
compute the quantity $\Ascr$, that is defined
in~\eqref{def:area-Poinc-sect}. The results are reported in the left
panel of Fig.~\ref{fig:confr_sec+compl}. By comparing that colour-code
plot with those included in Fig.~\ref{maxecc_maxdeltaomega}, we can
appreciate that there is good agreement between them: the most robust
regions (according to the criterion of the minimal area) look also
well apart from possible collisions (because the eccentricity of the
outer planet does not reach large values) and fairly inside the
librational regime with respect to the difference of the pericentre
arguments. The initial values of the mean anomalies that are expected
to correspond to the most robust orbit are the following ones:
\begin{equation}
  \label{frm:initial-mean-anomalies}
  M_1(0)=355^\circ\ ,
  \qquad
  M_2(0)=335^\circ\ .
\end{equation}
In the right panel of Fig.~\ref{fig:confr_sec+compl}, we have plotted
the intersections of the corresponding ``most robust'' orbit with
respect to the Poincar\'e hypersurface $\eta_2=0$. Moreover, we have
done the same also for the flow of the Hamiltonian $H^{({\rm sec})}$,
which is defined in~\eqref{frm:hsec}, starting from the corresponding
initial conditions $\big(\vet{\xi}(0),\vet{\eta}(0)\big)$ that are
computed in terms of the secular canonical coordinates.  The
comparison of these two different kinds of Poincar\'e sections allows
us to conclude that, for what concerns the most robust orbit, the
behaviour of the eccentricities in the case of the complete planetary
Hamiltonian should be rather close to that we can observe in the
secular model at order two in the masses.

\begin{figure}
\begin{center}
 \includegraphics[width=12.4cm]{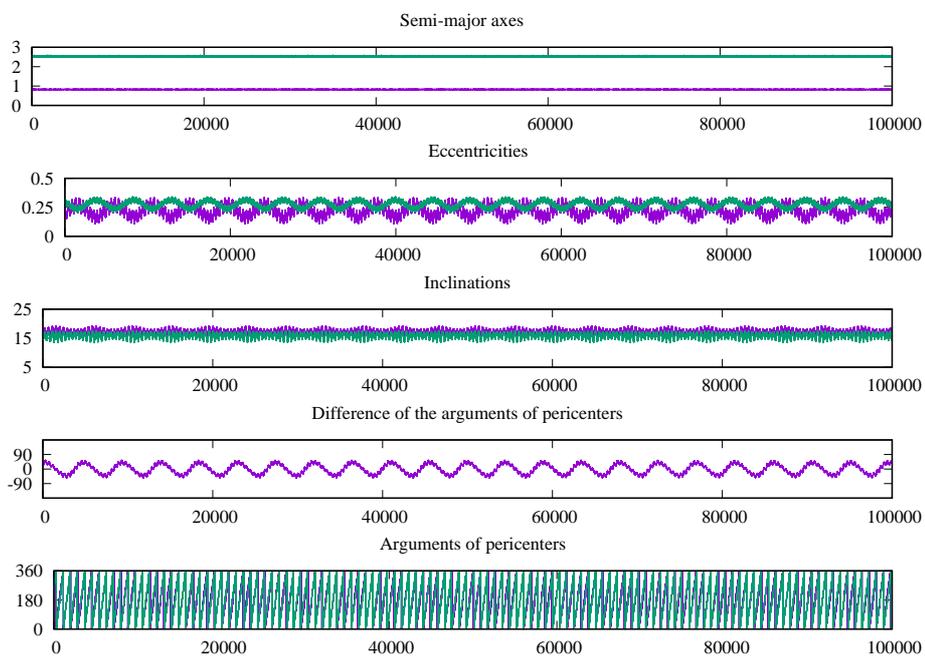}
 \caption{Orbital evolution of the exoplanets $\upsilon$~And~\emph{c}
   and $\upsilon$~And~\emph{d} in the case of the single set of
   initial conditions which is described in the Table included on the
   right of Fig.~\ref{fig:crit_min_area}.}
   \label{orb_scelti}
\end{center}
\end{figure}

Fig.~\ref{orb_scelti} describes the dynamical evolution of the
exoplanets $\upsilon$~And~\emph{c} and $\upsilon$~And~\emph{d} in the
case of the orbit that we consider as the most robust, according to the
analysis we have widely discussed in the present section.  The
comparison with the corresponding graphs that are reported in
Fig.~\ref{orbU} allows us to appreciate that the behaviour has now
become pleasantly quasi-periodic. Of course, this difference is
entirely due to our accurate choice of the initial conditions.

\section{Conclusions and perspectives}
\label{sec:conclu}
At the very beginning, our main motivation to start the investigations
we have described in the present paper was essentially of mathematical
character. Indeed, our aim was to select a set of initial conditions
corresponding to an invariant KAM torus whose existence could have
been proved rigorously. For such a purpose, the adoption of an
approach based on a Computer-Assisted Proof (hereafter, CAP) is
somehow unavoidable. In the last few years, the performances of CAPs
have been improved so much that they are able to prove the existence
of invariant tori for values of a small parameter (say, $\epsilon$)
that are amazingly close to the so called breakdown threshold, i.e.,
the critical value of $\epsilon$ beyond which the KAM manifold under
study disappears (see~\cite{Fig-Har-Luq-2017}).  For the time being,
so-successful results have been obtained for benchmark systems
(mappings with or without additional dissipative terms) that are quite
interesting but intrinsically simple. On the other hand, the
application of CAPs to realistic models of physical interest
highlights that there is still a gap to fill in order to approach the
numerical threshold (see, e.g., \cite{Cal-Cel-Gim-Lla-2022}
and~\cite{Val-Loc-2022}; see also~\cite{Car-Loc-2020} for the rigorous
evaluation of an effective stability time, with a similar kind of CAP
technique). This is the reason for which we were looking for initial
conditions that were not only corresponding to an invariant torus
(that could have been found by applying, e.g., the frequency analysis;
see~\cite{Laskar-03}), but also quite far from its breakdown threshold
(which is somehow depending on the physical parameters characterising
a planetary systems). This has been made with the hope that a rigorous
proof of the existence of such a KAM manifold would have been so
relatively easy to be completed even if the CAP technique we adopted
needs further improvements, to be extensively applied to Hamiltonian
models of physical interest.  This strategy of ours has been
successful: as it is discussed in~\cite{Car-Loc-San-Vol-2022}, in the
case of the secular dynamics of the $\upsilon$~Andromed{\ae} planetary
system we have been able to rigorously prove the existence of a KAM
torus that is travelled by the motion law starting from the initial
conditions we have selected and reported in the Table included on the
right of Fig.~\ref{fig:crit_min_area}.

In order to solve such a challenging problem, we have introduced a
robustness criterion that we have named ``of the minimal area''.  The
practical implementation of this method of investigation is
computationally inexpensive, making it suitable for extensive studies
of extrasolar systems. Indeed, it just requires a few additional
computations during the numerical integrations of the Hamilton
equations, each of them starting from different initial conditions,
that all together should give a reasonable coverage of a data range
which is compatible with the observations. Our robustness criterion is
also flexible enough to be applied jointly with numerical integrations
of a complete planetary model or a secular one without any need of
additional efforts for the adaptation. Moreover, the comparisons
reported in the right panel of Fig.~\ref{fig:confr_sec+compl} shows
that in the case of the selected initial conditions there is a good
agreement between the Poincar\'e sections for the secular model at
order two in the masses and those related to the complete planetary
system. Since the fraction of the chaotic motions is expected to be
much more relevant in the latter case than in the former one
(according to the discussions and figures widely commented in
Sections~\ref{sec:Ups-Andromedae_overview}--\ref{sec:Ups-Andromedae_sec_dyn}),
this result is not a priori obvious and enforces our confidence in the
accuracy of the secular model, at least in the region where the
invariant tori are more robust.

In our opinion, the possible applications of our approach are not
limited to problems which are interesting for reasons that are mainly
mathematical. From an astronomical point of view, we think that the
most interesting result described in this work of ours concerns with
the masses of the planets in the $\upsilon$~Andromed{\ae} system.  Our
analysis allow to conclude that configurations with a large mass of
$\upsilon$~And~\emph{c} have to be considered as more probable,
because they are more robust In other words, one can expect that
configurations with larger values of the mass of
$\upsilon$~And~\emph{c} are within a region that is extremely stable
because it is filled by tori so robust that they can eventually
persist also when other perturbing terms are considered. For instance,
additional gravitational effects could be taken into account, because
of the eventual reintroduction of $\upsilon$~And~\emph{b} and
$\upsilon$~And~\emph{e} in the planetary model.

The conclusion we have commented just above could be thought as
counter-intuitive, because one might expect that stability is always
gained by decreasing the values of the planetary masses. On the other
hand, the following easy remark could explain such a situation which
appears in contradiction: for fixed values of the semi-major axes and
the eccentricities, in the case of $\upsilon$~Andromed{\ae} system,
the configuration that we identify as the most robust among the
possible ones is that reducing as much as possible the imbalance
between the angular momenta\footnote{The angular momenta of
  $\upsilon$~And~\emph{c} and $\upsilon$~And~\emph{d} are such that
  $\Gscr_j=\Lambda_j\big(1-\sqrt{1-e_j^2}\big)$, with $\Lambda_j =
  {m_0 m_j\sqrt{G(m_0+m_j)a_j}}\big/{(m_0 + m_j)}$
  $\forall\>j=1,2$. Looking at the data about semi-major axes,
  eccentricities and minimal masses that are reported in
  Table~\ref{tab:orbel-partial}, one can easily check that the minimum
  difference between the angular momenta (i.e., $\Gscr_2-\Gscr_1\,$)
  corresponds to the maximum possible value of the mass of the inner
  planet and the minimum of that of the outer one, which are $m_1=1.91
  / \sin(\iota_1(0))$ and $m_2=4.22 / \sin(\iota_2(0))$, respectively,
  where the ranges of values of the initial inclinations $\iota_1(0)$
  and $\iota_2(0)$ are reported in Table~\ref{tab:orbel-with-errors}.}
of $\upsilon$~And~\emph{c} and $\upsilon$~And~\emph{d}. It is natural
to argue about the real meaning of such a possible explanation: is
this just by chance or is it quite general that planetary stability is
gained by a better balance of the angular momenta? If the latter
statement holds true, under which conditions?  We think that there are
also other natural questions about the generality of our approach,
which are mainly due to the fact that our robustness criterion has
been devised by studying the secular dynamics of a planetary
three-body problem in an apsidal locking regime. Could it be extended
to systems where the difference of the arguments of the pericentres is
in rotation? Could our approach be significantly adapted to systems
hosting more than two exoplanets? In our opinion, all these questions
deserve to be further investigated.

\vskip 1.2truecm
\noindent
{\bf Acknowledgements} This work was partially supported by the
MIUR-PRIN project 20178CJA2B -- `New Frontiers of Celestial Mechanics:
theory and Applications'.  The authors acknowledge also INdAM-GNFM and
the MIUR Excellence Department Project awarded to the Department of
Mathematics of the University of Rome `Tor Vergata' (CUP
E83C18000100006).

\end{document}